\documentclass[11pt,twocolumn]{aastex63}

\received{December 03, 2023}
\accepted{March 13, 2024}
\newcommand{\pmra}{$\mu_\alpha \cos\delta$}
\newcommand{\pmdec}{$\mu_\delta$}

\newcommand{\Teff}{T_\mathrm{eff}}

\begin{document}
\title{The Present-Day Mass Function of Star Clusters in the Solar Neighborhood}

\shorttitle{PDMF of star clusters in the Solar Neighborhood}

\shortauthors{Pang et al.}

\author[0000-0003-3389-2263]{Xiaoying Pang}
    \affiliation{Department of Physics, Xi'an Jiaotong-Liverpool University, 111 Ren’ai Road, Dushu Lake Science and Education Innovation District, Suzhou 215123, Jiangsu Province, P.R. China.}
    \email{Xiaoying.Pang@xjtlu.edu.cn}
    \affiliation{Shanghai Key Laboratory for Astrophysics, Shanghai Normal University, 
                100 Guilin Road, Shanghai 200234, P.R. China}
    
\author[0009-0002-8135-7011]{Siqi Liao}
    \affiliation{Department of Physics, Xi'an Jiaotong-Liverpool University, 111 Ren’ai Road, Dushu Lake Science and Education Innovation District, Suzhou 215123, Jiangsu Province, P.R. China.}

\author[0000-0002-3651-5482]{Jiadong Li}
    \affiliation{Max-Planck-Institut für Astronomie, Königstuhl 17, D-69117 Heidelberg, Germany}

\author[0000-0001-7395-1198]{Zhiqiang Yan}
    \affiliation{School of Astronomy and Space Science, Nanjing University, Nanjing 210093, P.R. China}
    \affiliation{Key Laboratory of Modern Astronomy and Astrophysics (Nanjing University), Ministry of Education, Nanjing 210093, P.R. China}

\author[0000-0002-5649-7461]{Mingjie Jian}
    \affiliation{Department of Astronomy, Stockholm University, AlbaNova University Center, Roslagstullsbacken 21, 114 21 Stockholm, Sweden}

\author[0000-0002-1805-0570]{M.B.N. Kouwenhoven}
     \affiliation{Department of Physics, Xi'an Jiaotong-Liverpool University, 111 Ren’ai Road, Dushu Lake Science and Education Innovation District, Suzhou 215123, Jiangsu Province, P.R. China.}

\author[0000-0003-4247-1401]{Shih-Yun Tang}
    \affiliation{Department of Physics and Astronomy, Rice University, 6100 Main Street, Houston, TX 77005, USA}
    \affiliation{Lowell Observatory, 1400 West Mars Hill Road, Flagstaff, AZ 86001, USA}

\author[0000-0003-4247-1401]{Yifan Wang}
    \affiliation{Department of Physics, Xi'an Jiaotong-Liverpool University, 111 Ren’ai Road, Dushu Lake Science and Education Innovation District, Suzhou 215123, Jiangsu Province, P.R. China.}

%-------------------------------------------------------------------------------------------% 
%*******************************************************************************************%
%TC:ignore
\begin{abstract} 
This work analyses the present-day mass function (PDMF) of 93~star clusters utilizing Gaia DR3 data, with membership determined by the \texttt{StarGo} machine learning algorithm. 
The impact of unresolved binary systems on mass estimation is rigorously assessed, adopting three mass ratio profiles for correction. The PDMF is characterized by the power-law index, $\alpha$, derived through a robust maximum likelihood method that avoids biases associated with data binning. 
The value of $\alpha$ for stars between the completeness limited mass of Gaia (with a mean 0.3\,$M_\odot$ for our cluster samples) and 2\,$M_\odot$, exhibits stability for clusters younger than 200\,Myr, decreasing for older clusters, particularly when considering stars within the half-mass radius. 
The PDMF of these star clusters is consistent with a dynamically evolved Kroupa IMF via the loss of low-mass stars. Cluster morphology shows a correlation with $\alpha$, as $\alpha$ values exhibit a decreasing trend from filamentary to tidal-tail clusters, mirroring the sequence of increasing cluster age. The dependence of $\alpha$ on total cluster mass is weak, with a subtle increase for { higher-mass} clusters, especially outside the half-mass radius.  We do not observe a correlation between $\alpha$ and the mean metallicity of the clusters. Younger clusters have lower metallicity compared to their older counterparts, which indicates that the older {  clusters} might {  have migrated} to the solar neighbourhood from the inner disk. A comparison with numerical models incorporating a black hole population suggests the need for observations of distant, older, massive open clusters to {  determine whether or not they contain black holes}.

\end{abstract}
%TC:endignore

%\keywords{stars: evolution --- open clusters and associations: individual -- stars: kinematics and dynamics -- methods: statistical -- methods: numerical }

%-------------------------------------------------------------------------------------------%
%*******************************************************************************************%
\section{Introduction}

The concept of initial mass function (IMF),  describing the distribution of stellar masses at birth, was first proposed by \citet{salpeter1955}. The IMF is mathematically formulated as a power-law function,
\begin{equation}\label{eq:IMF}
    \xi(m)=A m^{-\alpha}\,dm \quad ,
\end{equation}
where $A$ is a normalization coefficient, $m$ is the stellar mass, and $\xi(m)$ is a function that is either the number of stars or the normalized fraction of stars within the mass range $m+dm$.  $\alpha$ is the power-law index of the mass function, and $\alpha=2.35$ is referred to as ``Salpeter slope''. 

The IMF holds significant importance in comprehending numerous astronomical phenomena, including the formation of the first stars \citep{Bromm2009}, the evolution and formation of galaxies \citep{calura2009,2021A&A...655A..19Y}, and the precise determination of the absolute star formation rate \citep{2018A&A...620A..39J,aoyama2023}. Furthermore, the IMF plays a vital role in the theoretical framework concerning star formation, representing the ultimate outcome of molecular cloud contraction and fragmentation processes \citep{krumholz2014}.

Accumulating evidence suggests a deviation from the ``Salpeter slope'' in the mass function of lower-mass stars in the Galactic disc \citep{Kroupa1993,kroupa2001,chabrier2003}. In addressing this, \citet{Kroupa1993} proposed an alternative IMF, characterized by a multi-segment power law. This ``Kroupa IMF'' exhibits a shallower power-law index at lower masses {(in contrast to the Salpeter value, which is obtained from higher mass stars).

Numerous endeavors have been undertaken to determine the IMF in the Milky Way. {However, the Galactic field star population represents a diverse assembly of stars characterized by different ages and metallicities. Prior examinations of the field IMF frequently limit their focus to samples of low-mass stars to mitigate evolutionary effects over the age of the universe \citep{Kroupa1993,li2023Natur}. An accurate star formation history estimation of our galaxy, which degenerates with the estimation of the IMF for intermediate-mass and massive stars, remains challenging \citep{2019A&A...624L...1M}.

Young star clusters are optimal targets for determining the IMF. They contain a statistically significant number of stars and share similar characteristics, such as distance, age, and chemical composition. 
Moreover, a significant portion of stars in the Galactic field have their origin in unbound stellar groups that quickly dissipate after gas expulsion \citep{lada2003}. The filamentary-type stellar group identified in \citet{pang2022b} is one of these kinds, which is formed in environments with low star formation efficiency \citep{kruijssen2012}. Analyzing the mass function of such groups holds the potential to enhance our understanding of the role of dissolution in the evolution of mass distribution in these young stellar groups.

A major limitation in studying the IMF of star clusters arises from their dynamical evolution over time.  This evolution leads to a systematic depletion of low-mass stars through evaporation, and the initial loss of certain massive stars induced by dynamical ejection resulting from binary encounters, and subsequently due to stellar evolution. This process results in a stellar mass function that changes with cluster age  \citep{portegies2010,vesperini1997,baumgardt_dynamical_2003}. Therefore, accounting for the dynamical evolution of star clusters becomes essential when attempting to infer the IMF of a cluster from its present-day mass function (PDMF). Nearly all open clusters exhibit mass functions that resemble, or are consistent with, a dynamically evolved Kroupa/Chabrier-type IMF \citep{bastian_universal_2010}. 

When correcting the PDMF of star clusters for evolution, to obtain the IMF, all the aforementioned factors introduce uncertainties. 
Therefore, many measurements are typically presented in the form of the PDMF, with authors not attempting to deduce the IMF to avoid introducing additional uncertainties.

To obtain an accurate description of the PDMF,  star clusters with reliable memberships and high-quality photometry observations are essential. A robust membership avoids significant field contamination, preserving the anticipated PDMF. Precise photometry enables the accurate derivation of individual stellar masses. 
There are several additional factors that we need to consider when deriving the PDMF. The presence of unresolved binary stars should be accounted for, to avoid the introduction of a slight shift in the observed PDMF peak toward higher masses \citep{lee_origin_2020}. Simultaneously, the observational completeness of low-mass stars affects PDMF. For example, the 21~mag limit in the $G$ band of Gaia \citep{gaia_collaboration_gaia_2022} introduces limitations in detecting the lowest mass stars in star clusters. In particular, discerning evaporated low-mass stars situated in the periphery or in the tidal tail of a cluster presents a considerable challenge. The mass segregation effect biases the measured stellar population in the inner radius to massive stars \citep{2004A&A...416..137G,allison2009,pang2013}.
Despite utilizing space-based observatories like Gaia, the pronounced crowding of stars in the central region of distant clusters remains a formidable challenge for the observation of faint stars \citep{baumgardt_evidence_2023}. 

To satisfy the aforementioned criteria, and avoid the crowding effect of distant clusters, this work we focuses on star clusters about 500\,pc of the solar neighborhood. We are motivated to explore variations in the PDMF among star clusters of different ages and environments, aiming to reveal the PDMF of various star cluster types, to examine the influence of dynamical evolution on the mass distribution, and to identify the key parameters that affect the mass distribution in star clusters. 

This paper is organized as follows. In Section~\ref{sec:gaia}, we introduce the dataset and membership of star clusters. The PDMF is derived in Section~\ref{sec:pdmf}. In Section~\ref{sec:BC}, we adopt three different mass ratio profiles and correct the observed stellar mass for binaries. The power-law index of the PDMF is obtained via the maximum likelihood method in Section~\ref{sec:MLE}. In Section~\ref{sec:max_mass}, we discuss the evolution of maximum stellar mass in star clusters. The dependence of the PDMF on cluster parameters is investigated in Section~\ref{sec:MF_depend}. In Section~\ref{sec:dynamical}, we present how dynamical evolution affects the PDMF. We then investigate the relation between PDMF and the cluster chemical abundance in Section~\ref{sec:abundance}.  In Section~\ref{sec:BH}, we compare the PDMF to numerical models including black holes. Finally, we provide a brief summary of our findings in Section~\ref{sec:sum}.

%%%%%%%%%%%%%%%%%%%%%%%%%%%%%%%%%%%%%%%%
%%%%%%%%%%%%%%%%%%%%%%%%%%%%%%%%%%%%%%%%
%%%%%%%%%%%
\section{Membership from Gaia data}\label{sec:gaia}

In this study, we investigate a total of 93 star clusters.
The identification of member stars of 92 of the target star clusters was carried out in \citet{pang2021a, pang2021b, pang2022a,pang2022c}, and \citet{li2021}. These studies established stringent criteria, requiring member stars to have parallax and photometric measurements with uncertainties within a 10\% threshold. 
The selection of star cluster members is executed using the machine learning algorithm \texttt{StarGo} \citep{yuan2018} based on Gaia EDR\,3 and DR\,3 data \citep{gaia2021}. Member stars are chosen with a contamination rate of at most 5\%, resulting in a corresponding membership probability of 95\%.
The precision of the photometry has been improved for sources with magnitudes fainter than $G=13$\,mag following the correction of G-band photometry in Gaia DR\,3 by \citet{riello2021}. To enhance the accuracy of photometric and kinematic data for each star, we cross-match the members with Gaia DR\,3 \citep{gaia_collaboration_gaia_2022}, and our analysis relies on the data from Gaia DR\,3. We identify members of one old open cluster NGC\,6991 \citep[$>1\,$Gyr,][]{Cantat-Gaudin2020} in this work. The membership of this cluster was determined with the same method described above, using Gaia DR\,3 data (detailed in Appendix~\ref{sec:NGC6991}). In total, we have 93 star clusters in our sample for PDMF analysis. The basic parameters for these 93 clusters are presented in Table~\ref{tab:PDMF}.

% \xy{Yifan, Add Table here:}
\startlongtable
\begin{deluxetable*}{L RRR R RRR R}
	\tablecaption{PDMF and related parameters for 93 star clusters.
	\label{tab:PDMF}}
% 	\tablewidth{0pt}
	\tabletypesize{\scriptsize}
    % \tabletypesize{\footnotesize}
	\tablehead{
	    \colhead{Cluster}                  & 
        \colhead{Age}                      &
		\colhead{$\alpha_{\rm PDMF}$}      & 
        \colhead{$M_{\rm cl}$}            &
		\colhead{$m_{\rm max}$}           &
        \colhead{$m_{\rm lower}$}           &
        \colhead{${\rm [M/H]_{Gaia(DR3)}}$}    &
		\colhead{${\rm [M/H]_{APOGEE}}$}     &
		\colhead{Type}  
		\\
		\colhead{}                       & 
		\colhead{(Myr)}               & 
		\colhead{}                    &
		\colhead{($M_\odot$)}          &
  	    \colhead{($M_\odot$)}          &
		\colhead{($M_\odot$)}          &
  	    \colhead{}          &
		\colhead{}          &
		\colhead{}         
		\\
	    % \cline{1-9}
	    \colhead{(1)}   & 
	    \colhead{(2)}   & \colhead{(3)} & 
	    \colhead{(4)}   & \colhead{(5)} &
	    \colhead{(6)}   & \colhead{(7)} & 
	    \colhead{(8)}   & \colhead{(9)}
		}
	\startdata
	%generate date 2023-Nov-29 17:17:23--
%=========latex table==========
\mathrm{Alessi\ 20} & 9 & 1.76\pm0.17& 241 & 9.2 & 0.29 & -0.16\pm0.18 &         & f1\\
\mathrm{Alessi\ 20\ gp1} & 12 & 1.93\pm0.18& 172 & 6.4 & 0.26 & -0.08\pm0.14 &         & f2\\
\mathrm{Alessi\ 20\ isl1} & 100 & 1.72\pm0.32& 97 & 3.5 & 0.39 & -0.06\pm0.10 &         &        \\
\mathrm{Alessi\ 24} & 88 & 2.09\pm0.35& 119 & 5.4 & 0.42 & -0.11\pm0.17 &         &        \\
\mathrm{Alessi\ 3} & 631 & 1.41\pm0.24& 124 & 2.7 & 0.32 & -0.10\pm0.02 &         & t\\
\mathrm{Alessi\ 5} & 52 & 1.95\pm0.23& 244 & 6.1 & 0.39 & -0.19\pm0.10 &         &        \\
\mathrm{Alessi\ 62} & 691 & 1.71\pm0.38& 143 & 2.5 & 0.58 & -0.02\pm0.13 &         &        \\
\mathrm{Alessi\ 9} & 265 & 1.41\pm0.31& 61 & 3.4 & 0.26 & -0.13\pm0.03 &         &        \\
\mathrm{ASCC\ 105} & 74 & 1.94\pm0.48& 67 & 3.1 & 0.48 & -0.02\pm0.10 &         & f1\\
\mathrm{ASCC\ 127} & 15 & 1.70\pm0.18& 183 & 4.5 & 0.26 & -0.05\pm0.17 &         & f1\\
\mathrm{ASCC\ 16} & 10 & 2.09\pm0.17& 241 & 11.5 & 0.28 &  & -0.11\pm0.04 &        \\
\mathrm{ASCC\ 19} & 8 & 1.91\pm0.18& 198 & 6.3 & 0.27 & -0.09\pm0.21 & -0.11\pm0.12 & f2\\
\mathrm{ASCC\ 32} & 25 & 1.51\pm0.15& 577 & 6.8 & 0.45 &                &                &        \\
\mathrm{ASCC\ 58} & 52 & 1.65\pm0.18& 254 & 6.5 & 0.35 & -0.11\pm0.19 &         & f2\\
\mathrm{BH\ 164} & 65 & 2.37\pm0.25& 194 & 5.9 & 0.40 & -0.20\pm0.17 &         &        \\
\mathrm{BH\ 99} & 81 & 2.28\pm0.14& 558 & 5.8 & 0.39 & -0.04\pm0.15 &         & f2\\
\mathrm{Blanco\ 1} & 100 & 2.15\pm0.11& 343 & 3.4 & 0.23 & -0.11\pm0.14 &         & t\\
\mathrm{Collinder\ 132\ gp1} & 25 & 1.32\pm0.28& 144 & 7.0 & 0.37 &                &                &        \\
\mathrm{Collinder\ 132\ gp2} & 25 & 1.76\pm0.16& 362 & 9.0 & 0.38 &                &                &        \\
\mathrm{Collinder\ 132\ gp3} & 25 & 1.45\pm0.27& 121 & 5.0 & 0.34 &                &                &        \\
\mathrm{Collinder\ 132\ gp4} & 25 & 1.16\pm0.45& 68 & 6.9 & 0.44 &                &                &        \\
\mathrm{Collinder\ 132\ gp5} & 50 & 1.87\pm0.43& 53 & 4.2 & 0.40 &                &                &        \\
\mathrm{Collinder\ 132\ gp6} & 100 & 2.01\pm0.53& 39 & 2.7 & 0.45 &                &                &        \\
\mathrm{Collinder\ 135} & 40 & 1.59\pm0.13& 253 & 4.0 & 0.23 & -0.16\pm0.16 &         & f2\\
\mathrm{Collinder\ 140} & 50 & 1.74\pm0.17& 179 & 6.7 & 0.25 & -0.12\pm0.16 &         & f2\\
\mathrm{Collinder\ 350} & 589 & 1.34\pm0.28& 149 & 2.8 & 0.46 & -0.03\pm0.11 &         & t\\
\mathrm{Collinder\ 69} & 13 & 1.86\pm0.12& 402 & 10.1 & 0.29 & -0.33\pm0.09 & -0.16\pm0.08 & f1\\
\mathrm{Coma\ Berenices} & 700 & 1.29\pm0.20& 101 & 2.4 & 0.21 & -0.06\pm0.07 & -0.05\pm0.11 & t\\
\mathrm{Group\ X} & 400 & 2.02\pm0.25& 99 & 2.4 & 0.27 & -0.01\pm0.26 & 0.03\pm0.03 & d\\
\mathrm{Gulliver\ 21} & 275 & 1.88\pm0.31& 83 & 3.4 & 0.52 & -0.14\pm0.31 &         &        \\
\mathrm{Gulliver\ 6} & 7 & 2.08\pm0.20& 168 & 4.1 & 0.30 &  & -0.11\pm0.09 & f1\\
\mathrm{Huluwa\ 1} & 12 & 1.90\pm0.08& 740 & 10.8 & 0.23 & -0.10\pm0.18 &         & f1\\
\mathrm{Huluwa\ 2} & 11 & 1.90\pm0.11& 473 & 8.7 & 0.25 & -0.15\pm0.15 &         & f1\\
\mathrm{Huluwa\ 3} & 10 & 1.87\pm0.12& 375 & 9.9 & 0.24 & -0.20\pm0.23 &         &        \\
\mathrm{Huluwa\ 4} & 10 & 1.93\pm0.15& 182 & 5.7 & 0.21 &  &         & f1\\
\mathrm{Huluwa\ 5} & 7 & 1.84\pm0.27& 58 & 3.2 & 0.20 &  &         & f1\\
\mathrm{IC\ 2391} & 50 & 1.58\pm0.26& 140 & 5.0 & 0.29 & -0.04\pm0.12 &         &        \\
\mathrm{IC\ 2602} & 45 & 1.79\pm0.15& 188 & 5.2 & 0.21 & -0.21\pm0.18 &         &        \\
\mathrm{IC\ 348} & 5 & 1.99\pm0.32& 142 & 4.6 & 0.41 & -0.22\pm0.15 & -0.30\pm0.13 &        \\
\mathrm{IC\ 4665} & 36 & 1.74\pm0.30& 158 & 5.8 & 0.40 & -0.10\pm0.13 &         &        \\
\mathrm{IC\ 4756} & 955 & 1.54\pm0.17& 508 & 2.3 & 0.50 & -0.16\pm0.14 &         & t\\
\mathrm{LP\ 2371} & 19 & 1.64\pm0.38& 80 & 6.5 & 0.34 & -0.22\pm0.14 & -0.04\pm0.07 &        \\
\mathrm{LP\ 2373} & 4 & 1.75\pm0.22& 89 & 2.5 & 0.23 &  &         & f1\\
\mathrm{LP\ 2373\ gp1} & 10 & 1.92\pm0.17& 187 & 4.2 & 0.27 & -0.21\pm0.23 & -0.09\pm0.04 & f1\\
\mathrm{LP\ 2373\ gp2} & 9 & 1.92\pm0.11& 542 & 8.6 & 0.27 & -0.18\pm0.13 & -0.14\pm0.17 & f1\\
\mathrm{LP\ 2373\ gp3} & 6 & 1.81\pm0.22& 111 & 9.3 & 0.24 &  &         & f1\\
\mathrm{LP\ 2373\ gp4} & 6 & 1.92\pm0.14& 296 & 6.2 & 0.25 & -0.60\pm0.17 & -0.21\pm0.21 & f2\\
\mathrm{LP\ 2383} & 50 & 2.02\pm0.17& 281 & 5.6 & 0.30 & -0.12\pm0.10 &         & f2\\
\mathrm{LP\ 2388} & 22 & 1.93\pm0.22& 149 & 6.5 & 0.31 & 0.13\pm0.15 &         &        \\
\mathrm{LP\ 2428} & 200 & 2.34\pm0.32& 111 & 2.6 & 0.43 & -0.08\pm0.15 &         &        \\
\mathrm{LP\ 2429} & 1150 & 2.07\pm0.26& 147 & 2.1 & 0.44 & -0.08\pm0.13 &         & t\\
\mathrm{LP\ 2439} & 25 & 1.99\pm0.18& 143 & 4.0 & 0.24 & -0.08\pm0.13 &         & f2\\
\mathrm{LP\ 2441} & 75 & 2.19\pm0.31& 187 & 5.4 & 0.50 & -0.03\pm0.14 &         & f2\\
\mathrm{LP\ 2442} & 14 & 1.81\pm0.10& 319 & 5.3 & 0.17 &  &         & f2\\
\mathrm{LP\ 2442\ gp1} & 8 & 1.37\pm0.12& 105 & 4.6 & 0.09 &  & -0.02\pm0.12 & f2\\
\mathrm{LP\ 2442\ gp2} & 8 & 1.48\pm0.09& 144 & 2.3 & 0.09 &  & -0.27\pm0.39 & f2\\
\mathrm{LP\ 2442\ gp3} & 8 & 1.33\pm0.15& 60 & 2.0 & 0.10 &  & -0.12\pm0.09 & f2\\
\mathrm{LP\ 2442\ gp4} & 8 & 1.61\pm0.12& 107 & 3.6 & 0.11 &  & -0.16\pm0.16 & f2\\
\mathrm{LP\ 2442\ gp5} & 8 & 1.53\pm0.15& 72 & 4.2 & 0.11 & -0.09\pm0.03 & -0.09\pm0.09 & f2\\
\mathrm{Mamajek\ 4} & 371 & 1.91\pm0.24& 281 & 2.7 & 0.49 & -0.02\pm0.08 &         & t\\
\mathrm{NGC\ 1901} & 850 & 1.44\pm0.25& 124 & 2.1 & 0.36 & -0.11\pm0.10 & -0.17\pm0.08 &        \\
\mathrm{NGC\ 1977} & 3 & 1.74\pm0.18& 107 & 2.5 & 0.21 &  &         &        \\
\mathrm{NGC\ 1980} & 5 & 2.04\pm0.09& 753 & 9.7 & 0.26 & -0.13\pm0.17 & -0.50\pm0.71 &        \\
\mathrm{NGC\ 2232} & 25 & 1.70\pm0.17& 205 & 4.5 & 0.27 & -0.10\pm0.12 &         & f1\\
\mathrm{NGC\ 2422} & 73 & 1.90\pm0.18& 480 & 5.6 & 0.47 & -0.10\pm0.17 &         &        \\
\mathrm{NGC\ 2451A} & 58 & 1.96\pm0.18& 182 & 5.6 & 0.23 &  &         &        \\
\mathrm{NGC\ 2451B} & 50 & 1.81\pm0.15& 276 & 6.8 & 0.28 & -0.03\pm0.06 &         & f2\\
\mathrm{NGC\ 2516} & 123 & 2.02\pm0.07& 1973 & 4.9 & 0.36 & -0.11\pm0.18 &         & t\\
\mathrm{NGC\ 2547} & 40 & 1.80\pm0.12& 303 & 4.8 & 0.23 & -0.16\pm0.26 &         & f2\\
\mathrm{NGC\ 3228} & 63 & 1.55\pm0.37& 84 & 4.0 & 0.39 & -0.23\pm0.14 &         &        \\
\mathrm{NGC\ 3532} & 398 & 1.71\pm0.06& 2211 & 3.2 & 0.39 & -0.06\pm0.14 &         & h\\
\mathrm{NGC\ 6405} & 79 & 2.36\pm0.20& 596 & 5.6 & 0.54 & -0.13\pm0.17 &         &        \\
\mathrm{NGC\ 6475} & 186 & 1.53\pm0.09& 1026 & 4.1 & 0.36 & -0.02\pm0.11 &         & h\\
\mathrm{NGC\ 6633} & 426 & 1.28\pm0.21& 337 & 3.1 & 0.47 & -0.17\pm0.15 &         &        \\
\mathrm{NGC\ 6774} & 2650 & -0.04\pm0.30& 152 & 1.6 & 0.41 & 0.08\pm0.13 & 0.11\pm0.05 &        \\
\mathrm{NGC\ 6991} & 1400 & 0.95\pm0.19& 246 & 2.0 & 0.40 &                &                &        \\
\mathrm{NGC\ 7058} & 80 & 1.84\pm0.27& 122 & 4.5 & 0.33 & -0.04\pm0.12 & -0.03\pm0.09 &        \\
\mathrm{NGC\ 7092} & 350 & 1.94\pm0.17& 193 & 3.1 & 0.27 & -0.13\pm0.15 &         & t\\
\mathrm{Pleiades} & 125 & 2.01\pm0.09& 743 & 3.8 & 0.28 & -0.05\pm0.13 & -0.04\pm0.09 & t\\
\mathrm{Praesepe} & 700 & 1.92\pm0.10& 601 & 2.6 & 0.31 & 0.11\pm0.10 & 0.12\pm0.17 & h\\
\mathrm{Roslund\ 5} & 97 & 1.96\pm0.28& 191 & 4.7 & 0.46 & -0.11\pm0.17 &         & f2\\
\mathrm{RSG\ 7} & 70 & 1.59\pm0.49& 67 & 5.7 & 0.44 & -0.07\pm0.14 &         &        \\
\mathrm{RSG\ 8} & 18 & 1.79\pm0.18& 343 & 9.1 & 0.37 & -0.10\pm0.15 &         & f2\\
\mathrm{Stephenson\ 1} & 46 & 2.05\pm0.15& 263 & 6.6 & 0.27 & -0.08\pm0.19 &         & f1\\
\mathrm{Stock\ 1} & 470 & 1.30\pm0.31& 136 & 2.8 & 0.41 & 0.07\pm0.08 &         &        \\
\mathrm{Stock\ 12} & 112 & 1.49\pm0.31& 122 & 3.9 & 0.41 & -0.04\pm0.07 &         &        \\
\mathrm{Stock\ 23} & 94 & 1.22\pm0.56& 106 & 5.1 & 0.61 & -0.10\pm0.15 &         & f1\\
\mathrm{UBC\ 19} & 7 & 2.55\pm0.56& 42 & 5.3 & 0.36 &  &         &        \\
\mathrm{UBC\ 31} & 12 & 2.15\pm0.25& 258 & 8.4 & 0.42 & -0.12\pm0.18 &         & f1\\
\mathrm{UBC\ 31\ gp1} & 12 & 2.39\pm0.65& 58 & 4.4 & 0.46 & -0.27\pm0.12 &         & f1\\
\mathrm{UBC\ 31\ gp2} & 10 & 2.03\pm0.28& 186 & 13.8 & 0.39 & -0.15\pm0.17 &         & f1\\
\mathrm{UBC\ 7} & 40 & 1.82\pm0.15& 192 & 7.0 & 0.21 & -0.17\pm0.19 &         & f2\\
\mathrm{UPK\ 82} & 81 & 1.91\pm0.53& 57 & 5.1 & 0.46 & -0.05\pm0.17 &         &        \\
	\enddata
	\begin{tablecomments}{
        The ages of the clusters are obtained from \citet{pang2021a,pang2021b,pang2022a,pang2022c}, and are derived from PARSEC isochrone fitting. The age of NGC\,6991 is obtained in this work (see Appendix~\ref{sec:NGC6991}). 
        $\alpha_{\rm PDMF}$ is the power-law index of the PDMF for stars with a mass between $m_{\rm lower}$ and $2~M_\odot$. 
	    $M_{\rm cl}$ is the observed total stellar mass above the completeness mass limit of a star cluster.
	    $m_{\rm max}$ is the observed maximum stellar mass in each cluster. $m_{\rm lower}$ is the completeness limited mass of each cluster from Gaia DR\,3 data.
	    Columns~7 and~8 are mean metallicities of target clusters derived from Gaia DR\,3 XP spectra \citep{li2023b} and APOGEE, respectively. The error is the dispersion of the metallicity of each cluster. The last column is the morphological type taken from \citet{pang2022a}. 
	    }
    \end{tablecomments}
\end{deluxetable*}

\section{Present-day Mass function}\label{sec:pdmf}

\subsection{Binary correction}\label{sec:BC}

To construct the PDMF of {  the target clusters, it is necessary to first} compute the masses of individual member stars. Previous studies \citep{pang2021a,pang2021b,pang2022a,pang2022c,li2021} have obtained the best-fitting isochrones for the 92 star clusters by assessing the locations of member stars in the color-magnitude diagram {  (CMD)}. The same procedure is carried out for NGC\,6991 in this work (Figure~\ref{fig:ngc6991} in Appendix~\ref{sec:NGC6991}). Previous studies directly used the isochrone provided by PARSEC. This time we enrich the data points in each isochrone by a factor of four through interpolation. Following previous studies  \citep{liu2019}, we search for the nearest neighbour of each observed star on the isochrone using the KD-tree method \citep{mcmillan2007}, and assign the stellar mass of the nearest neighbor to the observed star as its stellar mass, $m_0$. 

{  The uncertainty in the stellar mass $m_0$ is primarily determined by the photometric uncertainty of Gaia DR3 \citep{gaia_collaboration_gaia_2022}, which may result in variations in the locations of cluster members in the CMD, and variations in the estimation of the binary properties \citep{pang2023}.
To determine the impact of these photometric uncertainties, we generate simulated observations. We assign each member a new magnitude drawn from a Gaussian distribution centered at the observed value and with a dispersion equal to the photometric uncertainty of the individual star. With the updated location of each member in the CMD, we repeat the aforementioned procedure, and assign the stellar mass, $m_0'$, that is equal the nearest point on the best-fitting isochrone of each cluster. This procedure is repeated 1000 times. The difference between the  $m_0$ and $m_0'$ is the adopted uncertainty of the stellar mass in this work. The average uncertainty in the stellar mass is 0.01 M$_\odot$. }

{  The} stellar mass, $m_0$, obtained in the manner described above is determined under the assumption that observed stars are single stars. {  However, as} shown by \citet{pang2023}, the unresolved binary fraction (for mass ratio $q>$0.4) among 85 star clusters ranges from 5\% up to 40\% \citep[Figure~2 (a) in][]{pang2023}. Note that these values are lower limits of the total unresolved binary fraction. Consequently, in order to reliably quantify the PDMF of {  the clusters in our sample}, we must correct individual stellar masses for binary effects.

\begin{figure}[tb!]
\centering
\includegraphics[angle=0, width=0.45\textwidth]{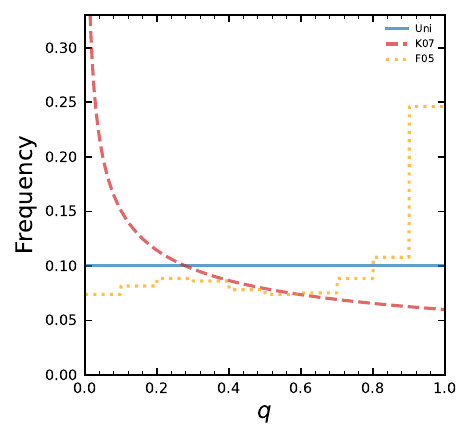}
    \caption{Distributions of mass ratio $q$ for binary systems. All three mass ratio distributions are normalized to unity. The uniform {  distribution (Uni)} is shown as the blue line, \citet[][K07]{kouwenhoven_primordial_2007} is shown as the red dashed curve, and \citet[][F05]{fisher_what_2005} as the orange dotted curve.
    }
\label{fig:mass_ratio}
\end{figure}

If an observed star is indeed an unresolved binary system, its total mass is under-estimated when using $m_0$. Our objective is to identify the corresponding primary and secondary masses for potential binary systems, which will be incorporated into the PDMF. \citet{lee_origin_2020} demonstrated that the influence of binarity on the mass function is sensitive to both the binary fraction and the mass ratio of the {  systems involved}.

To achieve this goal, we implement three mass ratio profiles in the binary correction, namely the uniform distribution, the profile of  \citet{kouwenhoven_primordial_2007}, and the profile of \citet{fisher_what_2005} (Figure~\ref{fig:mass_ratio}). A binary population is {  then} constructed based on each mass ratio distribution and the best-fitted isochrone for each cluster. Each point {  on} the isochrone is treated {  as a} single star, with its corresponding mass $m'$ designated as the primary mass. A mass ratio $q$ is randomly generated {  from one of the three} the mass ratio profiles. {  A companion star of mass} $q m'$, is assigned from the isochrone.  This process is repeated 100 times. For the low-mass primaries approaching the limit of 0.08\,$M_\odot$, their secondary mass $q m'$ is not available in the isochrone when $q$ is small. Consequently, the adopted $q$ values for these low-mass stars are preferentially higher than what is expected from the three adopted mass ratio distributions \citep[see][for details]{kouwenhoven2010}. 

We combine the luminosity of both primary and secondary stars in the $BP$, $RP$, and $G$ bands to compute the magnitude for a synthetic unresolved binary system in these {  filter bands}. Three populations of synthetic unresolved binary systems are generated following the aforementioned three mass ratio distributions. Given that the mean completeness mass limit of Gaia data for our cluster samples is 0.3\,$M_\odot$, exceeding 0.08\,$M_\odot$, the uncertainty associated with low-mass synthetic unresolved binaries featuring a primary component of $m'\sim0.08\,M_\odot$, does not impact on our findings (see Section~\ref{sec:MLE}).

\begin{figure}[tb!]
\centering
\includegraphics[angle=0, width=0.45\textwidth]{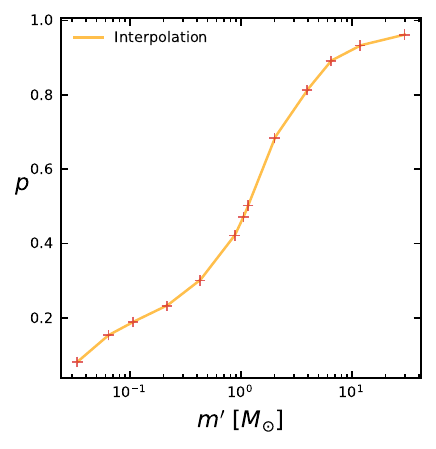}
    \caption{Dependence of binary probability $p$ on primary mass $m'$. The red crosses are empirical values obtained from \citet{offner_origin_2022}. The orange curve is the function that is interpolated by the red crosses.
    }
\label{fig:frequency}
\end{figure}

We assign a binarity probability $p$ for each observed star based on the empirical relation between binarity frequency and primary mass $m'$ (the crosses in Figure~\ref{fig:frequency}) taken from \citet{offner_origin_2022}.  
When adopting this relation, we assume that the correlation in question remains unaltered across varying cluster ages and the potential influence of dynamical evolution is neglected. We note that a higher initial binary fraction for low-mass stars has been suggested which decreases over time to the present-day binary fraction due to dynamical interactions \citep{1995MNRAS.277.1491K,1995MNRAS.277.1507K,2015ApJ...800...72T}. However, this dynamical evolution primarily results in the disruption of wide binary systems, while the impact on the hard binary systems, which are unresolved in our observations, is negligible. Given the substantial discrepancies in the observed and simulated dependencies of $p$ on $m'$, we consider the adoption from  \citet{offner_origin_2022} as an adequate choice.

By assuming the stellar mass $m_0$ of the observed stars as the primary mass, we interpolate the value of $p$ for an observed star based on this relation (orange curve in Figure~\ref{fig:frequency}). Utilizing this probability, each observed star is assigned either as an unresolved binary system or a single star. In the case of an unresolved binary system, we identify the synthetic unresolved binary system from the binary population, whose position is closest to the observed star in the {  CMD}. The corresponding primary and secondary masses of the binary system are considered in computing the PDMF (instead of $m_0$). This process is repeated 1000 times, during which the observed stars are treated as unresolved binary systems in $p \times 1000$ instances, and as single stars in $(1-p) \times 1000$ instances. Each process involves the determination of the PDMF.

\subsection{Power-law index $\alpha$ of the PDMF}\label{sec:MLE}

{  The Kroupa IMF \citep{kroupa2001} has a power-law index equal to 2.3 for $m>0.5$\,$M_\odot$ \citep[see section~3.2 in][]{yan2023}, when the influence of stellar density and metallicity on the IMF for initial cluster conditions is neglected \citep{jerabkov2018}. 
When determining the shape of the PDMF over a small stellar mass range, it is in many cases practical to adopt a single power law. This approach avoids the complications of the multiple parameters (the break point, and the two power indices) of a PDMF with two power law segments. This approach is commonly applied in star counting studies with a stellar mass detection limit similar to our completeness limit 0.3\,$M_\odot$ \citep{geha2013,gennaro2018,yang2024}.  
Therefore, we model the PDMF for our target clusters as a single power-law profile for the masses above the typical completeness limit 0.3\,$M_\odot$ in our sample. }

To determine the PDMF, we need to fit two parameters in Equation~\ref{eq:IMF}: the normalization parameter $A$, and the power-law index $\alpha$.  These parameters are sensitive to the approach how the data are binned \citep{maiz2005}.
To mitigate the bias arising from data binning, we adopt the maximum likelihood method to derive these coefficients.

To establish the likelihood function for a group of stellar masses, we assume that each mass is independently measured. Additionally, we postulate that the probability of obtaining a specific mass measurement is the mass function's value at that particular mass in  Equation~\ref{eq:IMF}. 

A sample of stellar masses of each cluster ($m_i$, where $i$ ranges from 1 to $N$, and $N$ represents the total number of members in each cluster) is taken after the binary correction. The likelihood function for this sample can be formulated as the product of the individual probabilities associated with each $m_i$:
\begin{equation}\label{eq: likelihood}
     \mathcal{L} = \prod^N_{i=1} p (m_i \mid \alpha, A) = \prod^N_{i=1} \xi (m_i, \alpha, A) \quad .
\end{equation}
Therefore, the log-likelihood of Equation~\ref{eq: likelihood} can be written as
 \begin{equation}\label{eq:loglikelihood}
     \log \mathcal{L} = \log A - \alpha \sum_{i=1}^N \log m_i \quad .
 \end{equation}
 Since $N$ is the integral of the PDMF from the lower stellar mass limit ($m_{\rm lower}$) to the upper stellar mass limit($m_{\rm upper}$),  
 \begin{equation}\label{eq: N}
     N = \int_{m_{\rm lower}}^{m_{\rm upper}} A m^{-\alpha} \mathrm{d}m = \frac{m_{\rm upper}^{1-\alpha} - m_{\rm lower}^{1-\alpha}}{1 - \alpha} A \quad .
 \end{equation}
 Here we express the parameter $A$ as a function of total member number $N$, $m_{\rm lower}$,  $m_{\rm upper}$ and power-law index $\alpha$ in Equation~(\ref{eq: N}). Given the incompleteness of the Gaia data, our member stars are subject to incompleteness beyond a specific magnitude. To avoid the bias introduced by incompleteness, we set the $m_{\rm lower}$ for each cluster as the completeness limited mass (dashed vertical red lines in Figure~\ref{fig:mass_function}), whose value is presented in column~6 in Table~\ref{tab:PDMF}. The maximum stellar mass in each cluster varies significantly. Old clusters lack high-mass stars as a result of stellar evolution. To ensure a consistent mass range {  in the analysis of} all target clusters, we only estimate the PDMF slopes for low-mass stars and adopt $m_{\rm upper}=2\,M_\odot$ (dotted vertical red lines in Figure~\ref{fig:mass_function}) for all clusters. 

% \jd{In this study, we employed the {\tt\string NumPyro}\footnote{https://num.pyro.ai} package to perform MCMC sampling. Specifically, we set the parameters $A$ and $\alpha$ with flat priors and conducted 1000 warm-up steps followed by 1000 fitting steps. Then the mean and the standard deviation of the resulting $\alpha$ values are given. }

In this work, we utilized the {\tt\string NumPyro} package\footnote{https://num.pyro.ai}, for performing Markov Chain Monte Carlo (MCMC) sampling. Flat priors are assigned to the parameters $A$ and $\alpha$, and a total of 1000 warm-up iterations are performed for each cluster, followed by 1000 sampling iterations. After this sampling process, the mean value of $\alpha$ is considered as the most probable power-law index of the PDMF between the completeness-limited mass to 2\,$M_\odot$ for each cluster. The mean standard deviation of $\alpha$ of all 1000 repeated processes is treated as its uncertainty. The value of the $\alpha$ and its uncertainty for each cluster is presented in column~3 of Table~\ref{tab:PDMF}.

% \tk{Note: in Figure 3 there are orange lines. But to me they look a bit yellow.}\xy{orange. }
 
In Figure~\ref{fig:mass_function} we present two examples of the PDMFs: Pleiades (left panels) and Praesepe (right panels). The most probable PDMFs (between completeness limit and 2\,$M_\odot$) are indicated with orange dashed lines. The histograms for the values of $\alpha$ values for all 93 target clusters are shown in Figure~\ref{fig:alpha_hist}. {  The peak at $\alpha\approx2.0$, agrees} with the value of the ``Kroupa IMF'' \citep{kroupa2001}, computed for the mass range of 0.3\,$M_\odot$ to 2\,$M_\odot$: $\alpha=2.04$. The distribution of $\alpha$ is nearly identical for the three mass ratio distributions adopted in the binary correction, indicating a negligible difference induced by the choice of the mass ratio profiles. Consequently, we only use the $\alpha$ value obtained from the uniform mass ratio distribution for subsequent analysis in the following sections. 

\begin{figure*}[tb!]
\centering
\includegraphics[angle=0, width=0.9\textwidth]{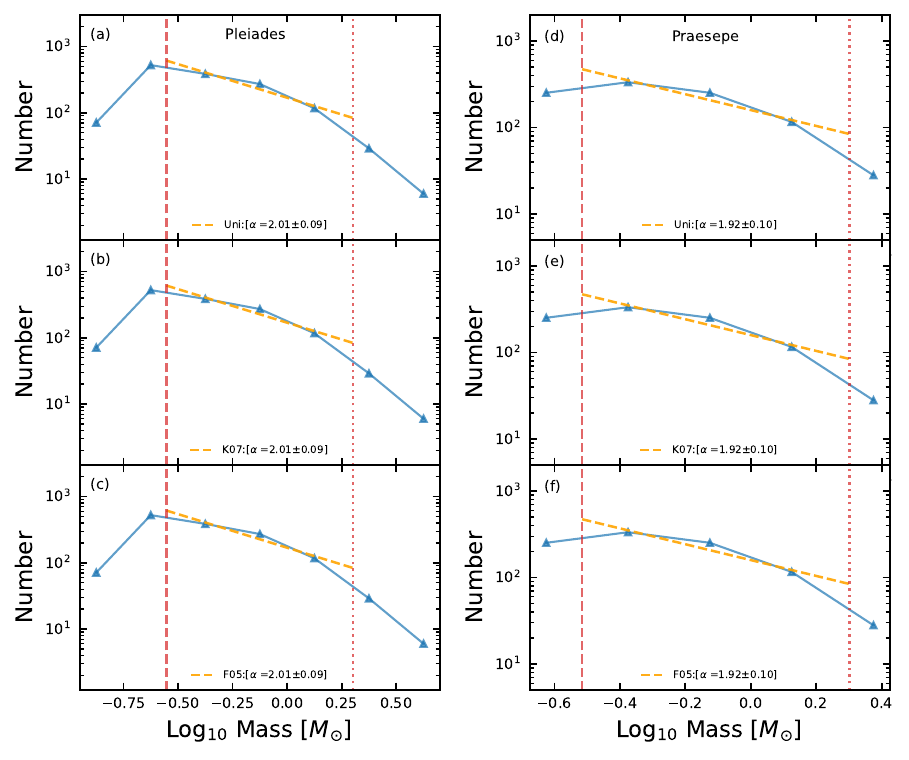}
    \caption{The most probable PDMFs of example clusters Pleiades (a, b, c) and Praesepe (d, e, f) after binary correction, considering three different mass ratio distributions. The red vertical dashed line indicates the completeness limited mass of Gaia DR\,3 data $m_{\rm lower}$, which is 0.28\,$M_\odot$ and 0.31\,$M_\odot$ for Pleiades and Praesepe respectively. The red vertical dotted line corresponds to the upper mass limit  $m_{\rm upper}$ of 2\,$M_\odot$. The blue curves are the mass distributions of Pleiades and Praesepe. The PDMFs (orange dashed lines) are determined only for the stellar mass between  $m_{\rm lower}$ and  $m_{\rm upper}$. The power-law index $\alpha$ value obtained from the maximum likelihood method in Section~\ref{sec:MLE} is indicated in each panel.
    }
\label{fig:mass_function}
\end{figure*}

\begin{figure}[tb!]
\centering
\includegraphics[angle=0, width=0.45\textwidth]{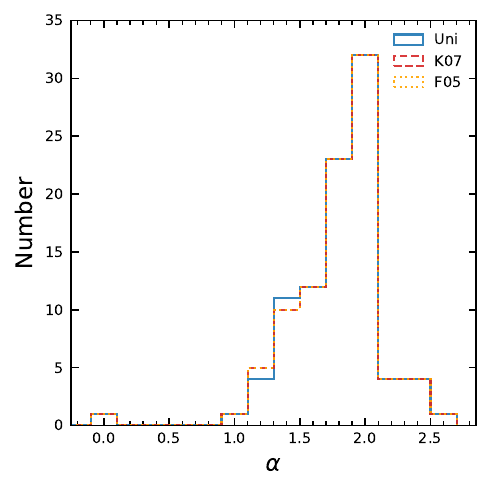}
    \caption{Histograms of the power-law indices ($\alpha$) of the PDMFs for a total of 93 star clusters. Histograms are shown for results obtained after binary corrections with different mass ratio distributions. The blue histogram shows the results obtained for a uniform (Uni) mass ratio distribution; the red dashed histogram for the mass ratio profile of \citet[][K07]{kouwenhoven_primordial_2007}; and the orange dotted histogram for that of \citet[][F05]{fisher_what_2005}. }
\label{fig:alpha_hist}
\end{figure}

\subsection{Maximum stellar mass in the cluster}\label{sec:max_mass}

In Figure~\ref{fig:max_mass} (a) we present the dependence of the observed maximum stellar mass, $m_{\rm max}$ (column~5 in Table~\ref{tab:PDMF}), on the observed total stellar mass above the completeness mass limit, $M_{\rm cl}$ (column~4 in Table~\ref{tab:PDMF}). Each data point represents one cluster, with its color indicating the cluster age. 
While the mass estimation method introduced in the previous sections is suitable for a stellar population, the uncertainties arising from dust extinction, binary probability, binary evolution, merger probability, and dynamical ejection become inherently irreducible through statistical arguments when assessing the mass of a single star. Addressing these factors typically yields a mass uncertainty exceeding 0.3 dex \citep{2023A&A...670A.151Y}. Conducting a detailed mass uncertainty analysis through dynamical simulations of star clusters is complicated and lies beyond the scope of the current study. Consequently, we adopt the largest $m_0$ as $m_{\rm max}$ without accompanying uncertainty estimation. The findings in this section should be interpreted as suggestive rather than conclusive.

As can be seen from Figure~\ref{fig:max_mass} (a), older clusters (orange to yellow dots) have a smaller maximum observed stellar mass, which can be primarily attributed to stellar evolution. {  This trend is more prominent in panel (b), which shows the relation between $m_{\rm max}$ and the cluster age.} A positive correlation is observed between $m_{\rm max}$ and $M_{\rm cl}$ (grey curve, {  panel (a)}), indicating that more massive young clusters generally host stars with a higher $m_{\rm max}$. This finding aligns with the observed $m_{\rm max}^{\rm ini}$--$M_{\rm ecl}$ relation reported in previous studies \citep{2010MNRAS.401..275W,yan_optimally_2017,2023A&A...670A.151Y}, in which $m_{\rm max}^{\rm ini}$ is the initial maximum stellar mass in a young star cluster, and $M_{\rm ecl}$ is the embedded cluster mass.

To further illustrate this, we compare the observed $m_{\rm max}$ values with the theoretical stellar maximum mass limit ({  thin} dashed lines). Adopting the stellar-mass--lifetime relation given by the PARSEC stellar evolution model version 1.2s \citep{bressan2012}, the theoretical stellar maximum mass limit is the most massive star that could exist at the age of the star cluster. We also plot the $m_{\rm max}^{\rm ini}$ value predicted by the $m_{\rm max}^{\rm ini}$--$M_{\rm ecl}$ relation ({  thick} dotted curve, using the GalIMF code, \citealt{yan_optimally_2017}). We note that the initial total stellar mass for the modeled embedded star cluster, $M_{\rm ecl}$, is transformed to $M_{\rm cl}$, the present-day mass above a completeness mass limit of $m_{\rm lower}=0.3~M_\odot$ (the mean completeness-limited mass in our cluster sample), to compare with our observations. The large scatter in the $m_{\rm lower}$ values in target clusters (Table~\ref{tab:PDMF}) introduces a large scatter of the data points in Figure~\ref{fig:max_mass} (a). In addition, the transformation from $M_{\rm ecl}$ to $M_{\rm cl}$ accounts only for the stellar evolution, which plays the predominant role during the phase of early mass loss in a cluster  \citep{2003MNRAS.340..227B}. The combined constraints are illustrated by the thin solid curves. This indicates that, in addition to the age of a star cluster, the total cluster mass also exerts an influential role in determining a stellar mass limit following the empirical tight $m_{\rm max}^{\rm ini}$--$M_{\rm ecl}$ relation.

\begin{figure*}[tb!]
\centering
\includegraphics[angle=0, width=1.0\textwidth]{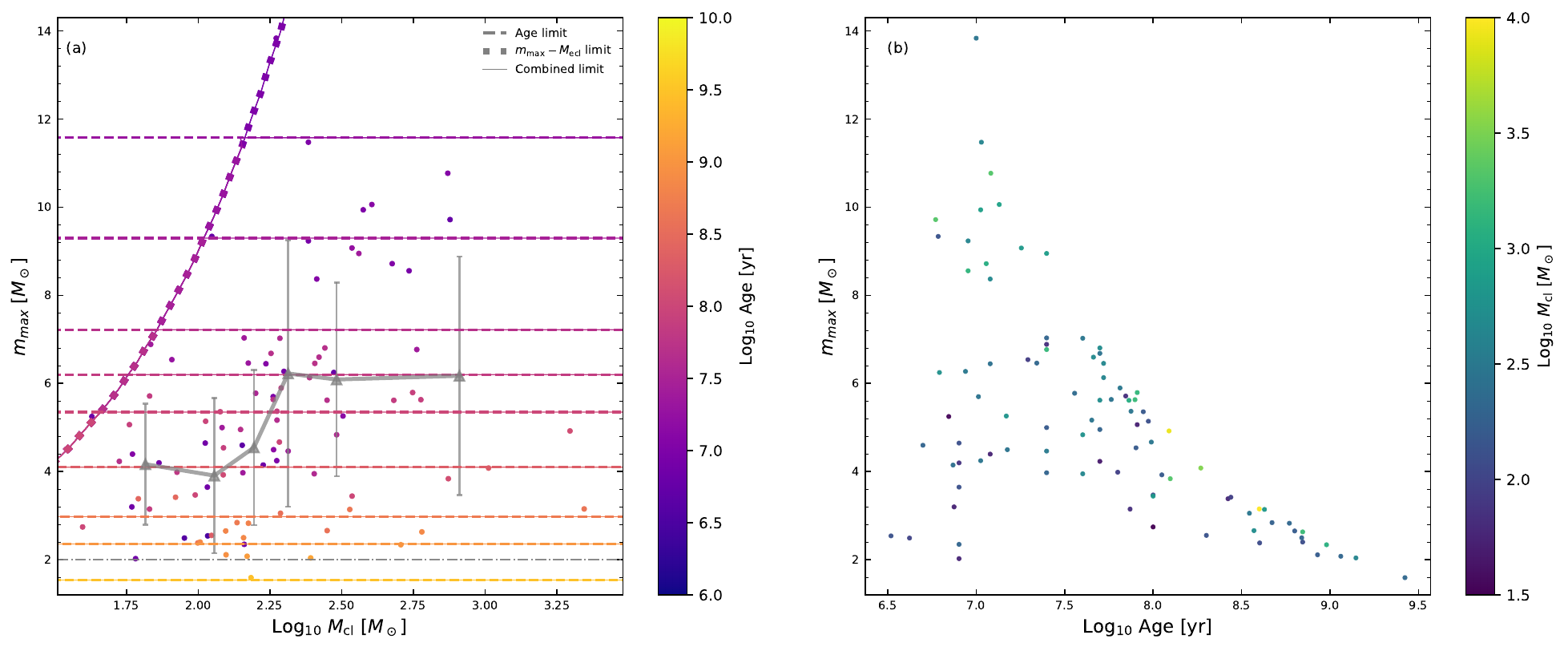}
    \caption{{  (a): Relation between $m_{\rm max}$, the maximum stellar mass in each cluster, and $M_{\rm cl}$, the total observed cluster mass above the completeness limit. Each dot represents one open cluster, and its color represents the cluster age, which is indicated by the color bar. The grey triangles in panel are average values of the total cluster mass and the maximum stellar mass for all 15 or 16~clusters in each bin, with the standard deviation indicated by the error bar. The horizontal grey dashed-dotted line indicates the upper mass limit, $m_{\rm upper}=2\,M_\odot$, adopted in the PDMF derivation. The thin dashed horizontal lines present the theoretical stellar maximum mass limit at a given age. The thick dotted curves indicate the values of $m_{\rm max}^{\rm ini}$ for embedded star clusters given by the relation $m_{\rm max}^{\rm ini}$--$M_{\rm ecl}$ \citep{yan_optimally_2017}. The combined constraints are illustrated by the thin solid curves. (b): Relation between $m_{\rm max}$ and the cluster age. The total cluster mass above the completeness limit, $M_{\rm cl}$, is indicated by the color bar.  }
    }
\label{fig:max_mass}
\end{figure*}

\section{Present day mass function dependencies  }\label{sec:MF_depend}

\subsection{On dynamical evolution}\label{sec:dynamical}

In Figure~\ref{fig:alpha_age_mass}, we display the distribution of the power-law index $\alpha$ of the PDMF along the cluster age (left panels); and along cluster total mass (right panels). We overplot the predicted $\alpha=2.04$ (grey dashed-dotted line) from the single power-law fit of the ``Kroupa IMF'' \citep{kroupa2001} in the mass range from 0.3\,$M_\odot$ to 2\,$M_\odot$. The value of $\alpha$ remains stable at $\alpha\approx 2$ for clusters younger than 200\,Myr, followed by a subsequent decline with increasing age (Spearman’s rank {  correlation coefficient  $s=-0.16$)}. This negative trend between $\alpha$ and cluster age becomes more pronounced among the population of stars within the half-mass radius $r_h$ (blue dots in panel (c)), {  with $s=-0.40$}. {  The half-mass radius $r_h$ is defined here as the radius of the sphere that contains half of the observed total mass of the cluster above completeness limit.} However, this trend is notably absent for the stars outside $r_h$. 
The deviation of the value of $\alpha$ from the “Kroupa IMF”, that is, the declining trend of $\alpha$ with age for stars within $r_h$, can be attributed to the dynamical evolution of the cluster \citep{baumgardt_dynamical_2003}. As clusters become older {  than $100-200$\,Myr}, which is a typical relaxation timescale of open clusters, an increasing number of low-mass stars escape through processes such as two-body relaxation, or being pulled away by the Galactic tidal field. Younger clusters have a higher $\alpha$ (steeper; a higher fraction of low-mass stars). As clusters age and approach dissolution, they experience a depletion in low-mass stars, resulting in a lower $\alpha$ (shallower; indicating a smaller fraction of low-mass stars). Nonetheless, the completeness-limited mass imposed by Gaia observations contributes, albeit partially, to the variability observed in the $\alpha$ values across clusters. Clusters with $m_{\rm lower}$ below the mean value of 0.3\,$M_\odot$ are all younger than 1\,Gyr, exhibiting a mean $\alpha$ of 1.80 that is consistent with the initial slope of the ``Kroupa IMF'' between 0.2 to 2\,$M_\odot$ with $\alpha_{0.2-2}^{\rm ini}=1.83$. On the other hand, clusters with $m_{\rm lower}$ exceeding 0.3\,$M_\odot$ are older on average, reaching 3\,Gyr in our sample. These older clusters have a mean $\alpha$ of 1.75 which is significantly lower than the slope of the ``Kroupa IMF'' in a similar mass range, being $\alpha_{0.4-2}^{\rm ini}=2.21$ from 0.4 to 2\,$M_\odot$. This confirms that the PDMF of older clusters has dynamically evolved and departed more significantly from the IMF slope than that of younger clusters. 

{  To determine the robustness of the declining trend of $\alpha$ with age, we derive the value of $\alpha$ for the PDMF of stars between the completeness limited mass and 1\,M$_\odot$. The correlation is identical to that of the PDMF in the left-hand panels of Figure~\ref{fig:alpha_age_mass}. This again confirms that dynamical evolution mostly affects low-mass stars, while the dynamics of high-mass stars is governed by rapid stellar evolution. The value of $\alpha$ decreases when relaxation becomes dominant (see Figure~\ref{fig:alpha_bh}). We do not obtain PDMF for stars above 1\,M$_\odot$ due to low-number statistics.}

A modest correlation between $\alpha$ and the total mass of clusters is observed. The power-law index $\alpha$ shows a slight increase toward more massive clusters, indicating that more massive clusters have a relatively larger fraction of low-mass stars within the selected mass range (completeness limit up to 2\,$M_\odot$). 
This positive correlation becomes slightly more significant for stars outside $r_h$ (orange dots), a Spearman’s rank correlation coefficient changing from $s=0.21$ to $s=0.28$. However, for stars inside $r_h$, no such dependence is observed ($s=0.11$). 

The dependence of $\alpha$ on total cluster mass 
indicates that more massive clusters contain a higher fraction of faint stars. In particular, most faint stars reside in the outskirts of the massive clusters. As the stellar population evolves dynamically, low-mass stars gain kinetic energy through two-body relaxation, causing them to migrate away from the cluster center, and inhabit the outer regions. Consequently, there are more low-mass stars outside the half-mass radius $r_h$. In low-mass clusters, the velocities of faint stars at the outskirt easily exceed the escape velocity of the cluster; these therefore quickly escape. Conversely, in more massive clusters, the deeper gravitational potential well can prevent the energetic low-mass stars in the periphery from escaping. Most young filamentary or fractal stellar groups are relatively low-mass. They are known to be in a state of disruption after gas expulsion and now are undergoing expansion \citep{pang2022a}. The dissolution state also speeds up the evaporation of low-mass stars in these young and low-mass stellar groups. This contributes to the overall observed trend in Figure~\ref{fig:alpha_age_mass} (d).

\begin{figure*}[tb!]
\centering
\includegraphics[angle=0, width=1.\textwidth]{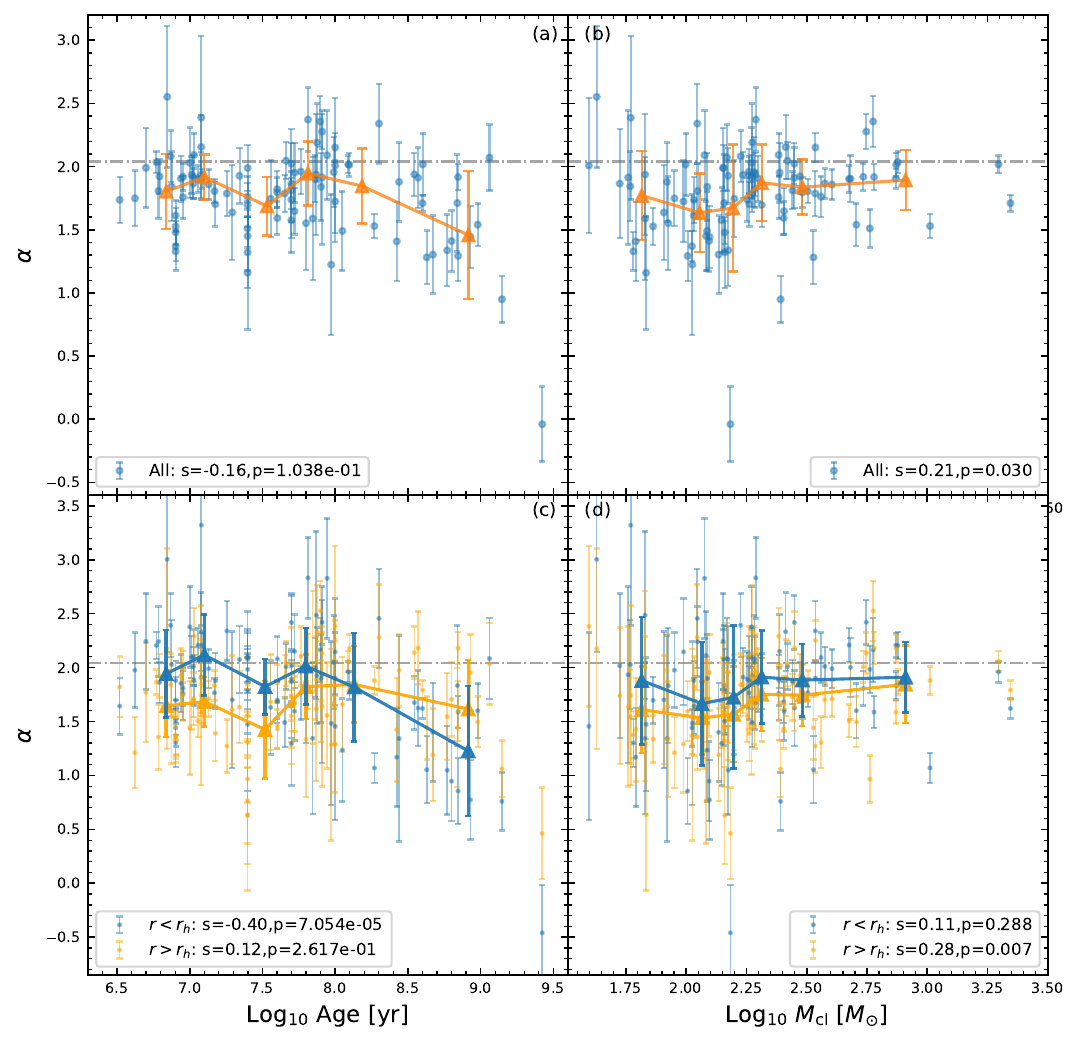}
    \caption{ Dependence of the power-law index ($\alpha$) of the most probable PDMF on the cluster age (left panels) and total cluster mass above completeness limit, $M_{\rm cl}$  (right panels). The uniform mass ratio distribution is used for binary correction. The grey dashed-dotted line in each panel corresponds to $\alpha=2.04$ computed from the \citet{kroupa2001} IMF in the mass range from 0.3\,$M_\odot$ to 2\,$M_\odot$. The values of $\alpha$ in panels (a) and (b) are computed for all members in the cluster. 
    The orange triangles are average values of cluster age and $\alpha$ for all 15--16 clusters in each bin, with the standard deviation indicated by the error bar. In panels (c) and (d), we separate each cluster into two parts and derive the PDMF individually: within the half-mass radius ($r<r_h$: blue dots) and outside half-mass radius ($r>r_h$: orange dots). The orange and blue triangles are computed in the same manner as in panels (a) and (b). We exclude the disrupted cluster Group~X from panels (c) and (d) since it is hard to define the cluster center for its two-piece-fragmented shape. 
    The quantity $s$ is Spearman’s rank correlation coefficient, and $p$ is the probability of the null hypothesis (i.e., that no correlation exists between two variables) of the correlation test. A $p$ value of less than 0.1 indicates that the null hypothesis is rejected. 
    }
\label{fig:alpha_age_mass}
\end{figure*}

\citet{pang2022a} categorized their 85~star clusters into four distinct morphological types: filamentary, fractal, halo, and tidal-tail. These morphological classifications represent a continuum of stellar density and cluster age \citep{pang2022a}. The distribution of 
$\alpha$ values for each cluster type is summarized in Figure~\ref{fig:alpha_cluster_type}. It is observed that $\alpha$ values tend to decrease from filamentary to tidal-tail clusters, exhibiting a considerable amount of scatters. Halo and tidal-tail clusters are generally older ($>100$\,Myr) compared to filamentary and fractal type clusters ($<100$\,Myr), aligning with the trend seen in the dependence of $\alpha$ on cluster age in Figure~\ref{fig:alpha_age_mass} (left panels). There are two outliers in the group of filamentary clusters: Stock\,23 and UBC\,31\,gp1. These two low-mass clusters have fewer than 100 stars available for the PDMF derivation within the completeness limited mass $m_{\rm lower}$ and $m_{\rm upper}=2\,M_\odot$, and therefore have larger uncertainties {  in} $\alpha$. 

\begin{figure}[tb!]
\centering
\includegraphics[angle=0, width=0.45\textwidth]{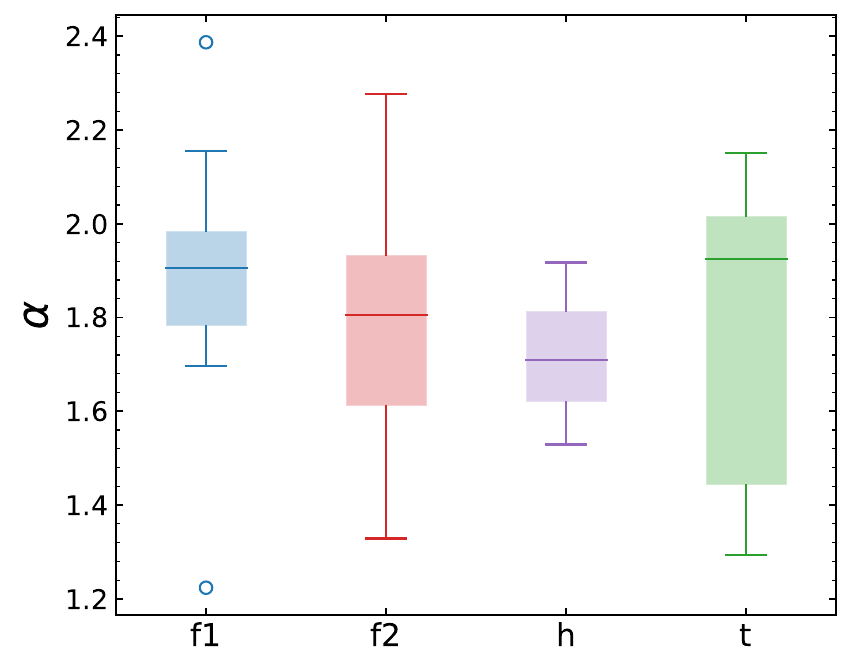}
    \caption{Dependence of the power-law index ($\alpha$) of the most probable PDMF on cluster morphologies: f1: filamentary (blue), f2: fractal (red), h: halo (purple), and t: tidal tail (green). The colored rectangles indicate the inner quartile range (IQR, 75th percentile minus 25th percentile). The median value is indicated with a horizontal line. The upper and lower whiskers of each colored rectangle mark the maximum and minimum values within 1.5 IQR. Clusters outside 1.5 IQR (whiskers) are outliers indicated as open circles: Stock\,23 and UBC\,31\,gp1 (blue circles).
    }
\label{fig:alpha_cluster_type}
\end{figure}

\subsection{On metallicity}\label{sec:abundance}

In this section, we explore the correlation between the PDMF and the metallicity ([M/H]) of star clusters. 
The metallicities for our sample stars are taken from \citet{li2023b}, who utilized a neural-network-based regression model AspGap to process Gaia DR\,3 XP spectra. 
This model was trained using APOGEE spectra data \citep{abdurrouf2022ApJS..259...35A}. 
We also cross-match the member stars in our target clusters with the data release~17 of APOGEE survey \citep{abdurrouf2022ApJS..259...35A} to obtain their chemical abundances. 
To ensure robust statistics, we consider only the clusters with a minimum of three stars that have measurable metallicity, allowing us to compute both the mean and dispersion of metallicity within each cluster.

To estimate the mean and standard deviation of the metallicity ([M/H]) for each cluster, we apply the Maximum Likelihood Estimation (MLE) method. 
We assume that the distribution of [M/H] values follows a Gaussian distribution for each cluster. 
The MLE technique enables us to identify the most probable mean ($\mu_{[\rm M/H]}$) and standard deviation ($\sigma_{\rm [M/H]}$) values for this distribution based on the observed data.

%\tk{In the paragraph below, what is the "number of samples per cluster"? Is it the number of stars with metallicity measurements? Or the number of members? The variable $N$ has already been used in the paper for other purposes, I think.}\xy{fixed, please check}

In constructing the likelihood function, we take into account the number of member stars in each cluster ($N$), the measured [M/H] values ($x_i$), and the individual measurement uncertainty ($\delta_i$). By maximizing this function, we can determine the values of the parameters that provide the optimal fit to the observed metallicity. We compute the mean [M/H] value and the inherent dispersion that maximizes this likelihood function:
\begin{equation}
    \mathcal{L} = \prod_{i=1}^{n} \frac{1}{\sqrt{2\pi (\sigma_{\rm [M/H]}^2 + \delta_{i}^2)}} \exp\left(-\frac{(x_i - \mu_{\rm [M/H]})^2}{(\sigma_{\rm [M/H]}^2 + \delta_{i}^2)}\right) \quad .
\end{equation}

We note that the metallicity measurements are subject to uncertainties, particularly for low-temperature stars. 
Commonly used stellar atmosphere models are optimized for FGK stars. Obtaining atmospheric properties of M-type stars is complex due to the presence of numerous broad absorption features from molecular lines that alter the atmosphere's structure \citep{Iyer2023}. This issue extends to the MARCS models utilized in the APOGEE survey. 
A distinct pattern is evident in the [M/H]--$\Teff$ diagram for the stars with $\Teff < 4500$\,K. Therefore, we exclude stars with $\Teff < 4500$\,K when establishing the relationship between the power-law index $\alpha$ and metallicity, to avoid large uncertainties. The measured values of [M/H] and the corresponding dispersions for the target clusters are presented in columns~7 and~8 in Table~\ref{tab:PDMF}.

In Figure~\ref{fig:alpha_M/H}, the dependencies of the PDMFs on metallicity obtained from stars with effective temperature $\Teff>4500$\,K are presented. No discernible trend between metallicity and $\alpha$ is seen in both Gaia DR\,3 data and APOGEE data. In contrast to both theoretical and observational investigations \citep{2008IAUS..255..285S,2015ApJ...806L..31M,2020ARA&A..58..577S,2020A&A...637A..68Y,2022MNRAS.509.1959S,li2023Natur} that propose a strong correlation between metallicity and the low-mass IMF slope, our sample does not provide evidence for such a relationship. This discrepancy may arise from the predominant similarity in metallicity across most of {  the clusters in our sample, resulting in negligible variations in the IMF}. In particular, the PDMF may not be sensitive to the expected low-mass IMF differences that might be washed out by the pronounced effects of dynamical evaporation.

An additional noteworthy pattern emerges, indicating that younger clusters (blue to magenta dots) exhibit lower metallicity compared to their older counterparts (orange to yellow dots). The older and more enriched clusters potentially might not have originated in the solar neighborhood; rather, they might have migrated from the inner disk to their current local positions through radial migration processes \citep{minchev_is_2009}. 
{  Recently, \citet{pang2022c} discovered the Collinder 132–Gulliver 21 Stream (270\,pc long) in the solar neighborhood, with stellar populations showing an age difference up to 250\,Myr. The oldest generation in the stream is approximately 0.3 dex (based on Gaia DR3 data) more metal-rich than the youngest generation. This may provide additional evidence of radial migration in the solar neighborhood. }
A similar pattern is also found in other open clusters at larger distances from the Sun (see, e.g., \citealt{Magrini2023}).

\begin{figure*}[tb!]
\centering
\includegraphics[angle=0, width=1.\textwidth]{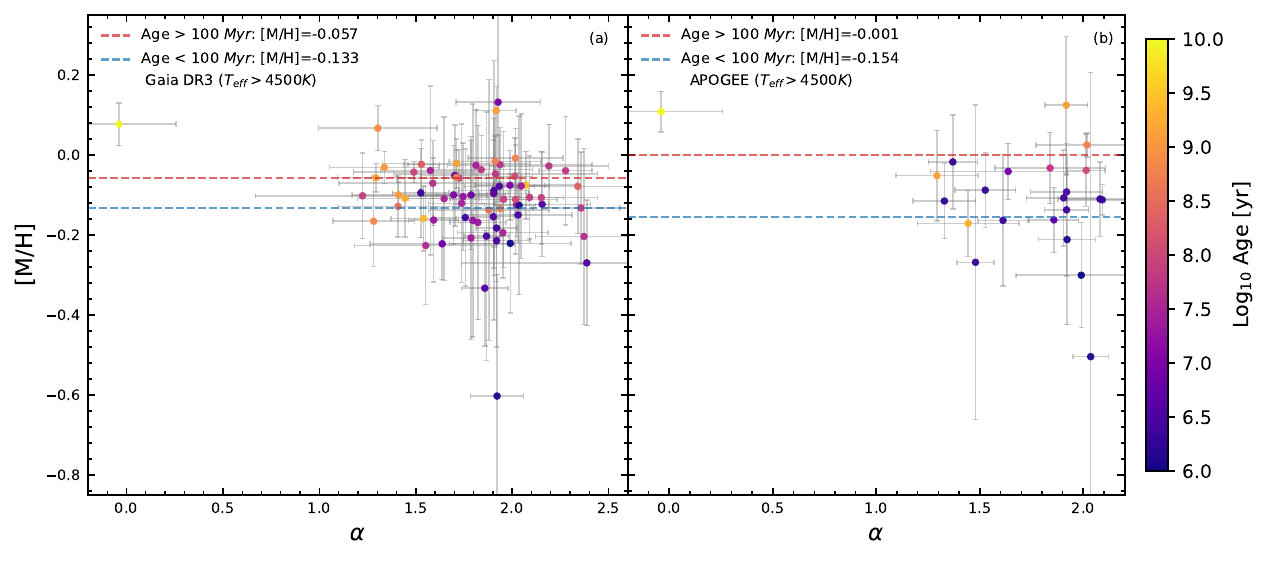}
    \caption{Distributions of the power-law index ($\alpha$) of the most probable PDMF and cluster metallicity [M/H]. The age of the cluster is indicated by the color bar on the right. The metallicity of clusters [M/H] in (a) and (b) is obtained using members that have [M/H] available from Gaia DR\,3 (a) and APOGEE (b), and that also have $\Teff>4500$\,K. Faint stars with $\Teff<4500$\,K have been excluded due to their large uncertainties. 
    }
\label{fig:alpha_M/H}
\end{figure*}

\subsection{On black holes}\label{sec:BH}

{   
\citet{torniamenti_stellar-mass_2023} proposed that stellar-mass black holes (BHs) can be retained even in open clusters with low escape velocities ($<3\,{\rm km\,s^{-1}}$). Their models incorporated 2–3 BHs in the Hyades cluster, leading to an increased central velocity dispersion of observed star clusters. However, distinguishing this BH signature from the effects of binary stars remains challenging. 
}

{   
We do not observe stars that are sufficiently massive to be candidate BH progenitors. Given that most clusters in our sample are older than 10\,Myr, BH progenitors should already have evolved. The most massive stars in the young ($<10$\,Myr) star clusters, on the other hand, are not massive enough to evolve into BHs. However, this cannot exclude the that BHs may have been present in the sample of young clusters in the past, because the massive BH progenitors can be ejected due to dynamical 3-body and 4-body interactions \citep{oh2015}. Even for the clusters that lack BH progenitors from stellar evolution, there is still the possibility to produce stellar-mass BH via binary mergers and frequent physical collisions \citep{oh2018,spera2019,dicarlo2020}. 
Consequently, we suggest at most a few BHs may exist in our sample of open clusters.  
}

{   
In globular clusters, on the other hand, stellar-mass BHs form a subsystems \citep{Breen2013}. When a BH subsystem contains more than 40 BHs, this system can significantly impact the cluster dynamics. 
\citet{baumgardt_evidence_2023} carried out star cluster simulations with retention of a BH subsystem by adopting low-velocity kicks. Their findings reveal that the presence of a BH subsystem significantly accelerates the two-body relaxation between BHs and massive stars. During the phase of fast relaxation, approximately 15 to 20 percent of the most massive stars are lost. Consequently, the retention of BHs in star clusters primarily amplifies the escape rate of higher-mass stars, and this effect can be observed through PDMF. 
}

We compare the power-law index $\alpha$ of the PDMF from our cluster sample with the models of \citet{baumgardt_evidence_2023} in Figure~\ref{fig:alpha_bh}. The models (plus signs) align with the lower edge of our cluster sample. Differentiation between various BH retention models primarily occurs in dynamically old clusters (age $>$ relaxation time). {  The relaxation time of each cluster is computed with equation~(7.108) in \citet{binney2008}, using the half-mass radius, the total mass, and the number of members in each cluster.} However, the majority of clusters in our sample are younger than one relaxation time, indicating a scarcity of old open clusters in the solar neighborhood for model comparison. 

The solar neighborhood, spanning a radius of 500\,pc, resides within the Local Arm \citep{reid2019} and is characterized by abundant young star formation regions and young star clusters. The older clusters in our sample likely did not form in situ, but instead underwent radial migration from the inner disk (as discussed in Section~\ref{sec:abundance}). Given that the typical survival timescale of open clusters is on the order of a hundred million years, the observed aged open clusters are skewed toward those that were initially massive at birth or were located in an orbit that experienced less frequent tidal interactions, such as those positioned in the direction of the Galactic anti-center. That is, old clusters are mostly to be found at larger distances. However, the PDMF of distant old open clusters is significantly influenced by the completeness mass limit of Gaia data. This should be properly addressed to facilitate meaningful comparisons with nearby clusters in the sample. 

Note that in \citet{baumgardt_evidence_2023}, a BH population comprises a few dozen BHs. The presence of this many BHs can significantly influence the evolution of the system. On the other hand, in open clusters \citep[which typically only contain a few black holes,][]{torniamenti_stellar-mass_2023},  {  the influence of these stellar-mass BHs on the PDMF is limited. Instead, the impact of these BHs is similar to that of massive compact binary systems in the core: relaxation due to gravitational interactions with neighbouring stars.} 
Additionally, due to the higher metallicity of open clusters compared to globular clusters, the masses of stellar-mass BHs formed in open clusters are considerably lower \citep{belczynski2010}. This further weakens the influence of BHs in open clusters. Finally, the influence of a stellar-mass BH on the stellar kinematics only becomes apparent on a relatively long time scale. The older clusters are therefore preferred targets to investigate the potential presence of BH candidates. Thus, distant, old, and massive open clusters could provide an opportunity to detect stellar-mass BHs.

\begin{figure}[tb!]
\centering
\includegraphics[angle=0, width=0.5\textwidth]{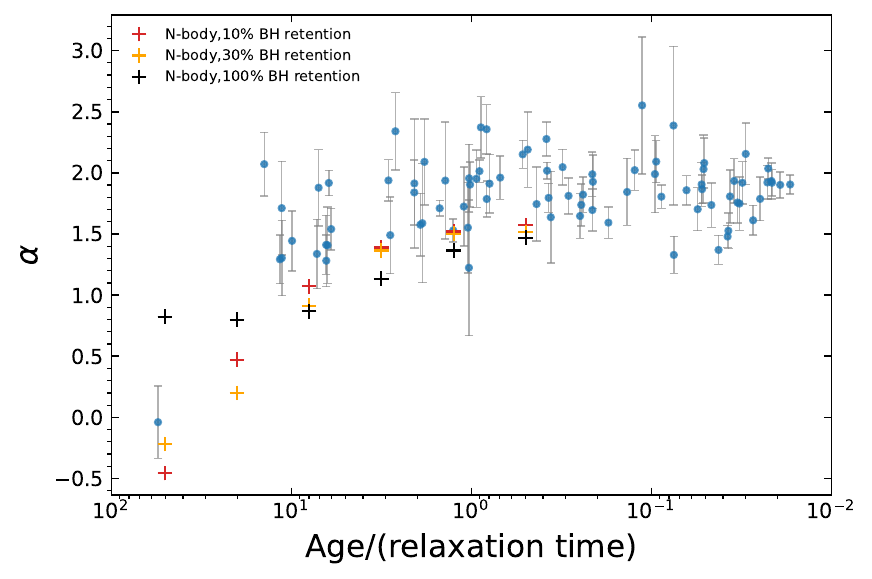}
    \caption{Dependence of the power-law index $\alpha$ of the most probable PDMF on the age, in units of relaxation time. Blue dots are 93~clusters in our sample. The plus symbols are taken from the $N$-body models of \citet{baumgardt_evidence_2023}. Red, orange, and black plus symbols represent models with a 10\%, a 30\%, and a 100\% BH retention rate, respectively.
    }
\label{fig:alpha_bh}
\end{figure}

\section{Summary}\label{sec:sum}

We analyze the PDMF of 93~star clusters using Gaia DR\,3 data. The stellar membership of {  92 of the} clusters is taken from previous studies \citep{pang2021a,pang2021b,pang2022a,pang2022c,li2021}. These studies determined stellar membership using the machine learning algorithm \texttt{StarGo}. {  Membership of the remaining cluster, NGC\,6991, is} identified in this work. 
We evaluate the impact of unresolved binary systems on stellar mass estimation. We adopt three mass ratio profiles: a uniform profile, the K07 profile \citep{kouwenhoven_primordial_2007}, and the F05 profile \citep{fisher_what_2005} for the correction of the observed stellar mass. A binarity probability is assigned to observed stars, based on the empirical relation between the binary frequency and primary mass taken from \citet{offner_origin_2022}. 

We characterize the PDMF with a power-law index, $\alpha$, which is determined using a maximum likelihood method that avoids biases introduced by data binning. The likelihood function is formulated for each cluster, allowing for robust statistical analysis. The value of $\alpha$ is only fitted for a given mass range: between the Gaia completeness limited mass (lower limit) and a consistent upper mass limit of 2\,$M_\odot$ for all clusters. The resulting histograms of $\alpha$ values, considering different mass ratio distributions for binary corrections, exhibit negligible differences.

We explore the relationship between the maximum observed stellar mass, $m_{\rm max}$, and the observed total mass above the completeness limit of the star clusters. The analysis reveals a positive correlation between $m_{\rm max}$ and the total cluster mass, where more massive young clusters tend to host stars with a higher $m_{\rm max}$. Older clusters currently have smaller $m_{\rm max}$ values due to stellar evolution. Our results suggest that, beyond the age of star clusters, their total mass plays a significant role in determining the maximum stellar mass limit, reinforcing previous findings in the literature \citep{2023A&A...670A.151Y}.

Based on the PDMF of these 93~star clusters, we investigate the dependence of the power-law index, $\alpha$, on dynamical evolution, metallicity, and black holes.

\begin{itemize}
\item There is a notable correlation between the power-law index, $\alpha$, of the PDMF and the age of the clusters. Specifically, the $\alpha$ value for clusters younger than 200\,Myr remains relatively stable at a value of $\alpha \approx 2$, which is consistent with the $\alpha$ value for a single power-law fit of the canonical \citet{kroupa2001} IMF in the mass range between 0.3 and $2~M_\odot$, but declines for older clusters. This negative trend becomes more pronounced when considering stars within the half-mass radius. The dynamical processes within the cluster, such as two-body relaxation and the Galactic tidal stripping, contribute to the removal of low-mass stars and produce the observed variation in  $\alpha$.

\item The dependence of the PDMF slope on the total cluster mass is weak, with a slight increase in the $\alpha$ value for more massive clusters. This dependence becomes slightly more pronounced when focusing on stars outside the half-mass radius. As low-mass stars migrate to the outer region of the cluster via two-body relaxation, the cluster potential determines whether they will remain bound to the cluster or whether they will escape. Less massive clusters exhibit a higher rate of escaping faint stars as a result of a lower escape velocity. 

\item Regarding cluster morphology, we find a decreasing trend in the values of $\alpha$ from filamentary to fractal, to halo and to tidal-tail clusters, which are also a sequence in terms of increasing cluster age. This resembles the correlation between the $\alpha$ value and the cluster age and again emphasizes the importance of secular dynamical evolution in shaping the PDMF of star clusters. 

\item Utilizing data from \citet{li2023b} and APOGEE survey, we examine the relation between the metallicity ([M/H]) of cluster members and the power-law index $\alpha$ of the PDMF. We exclude stars with $\Teff < 4500K$\,K due to uncertainties in metallicity measurements for these stars. Our results reveal no discernible trend between the value of $\alpha$ and the mean metallicity derived from stars with $\Teff > 4500K$. However, a distinct pattern emerges: younger clusters exhibit a lower metallicity compared to their older counterparts. This discrepancy is attributed to potential radial migration, indicating that the older, more metal-rich clusters may have originated from the inner circle of the solar neighborhood and subsequently migrated outward.

\item We attempt to explore the role of BHs in shaping star cluster PDMF, building on the findings of recent studies. We study the evolution of the $\alpha$ value with age (measured in relaxation times), and make a comparison between our observational results with the numerical simulations by \citet{baumgardt_evidence_2023} that included a BH population in their star cluster models. The differentiation between BH retention models is more pronounced in dynamically old clusters, whereas the majority of the analyzed star clusters are younger than one relaxation time. Therefore, the current star cluster sample cannot provide strong constraints for these models. There is a need for more distant, older, and more massive open clusters to explore the potential presence and impact of stellar mass BHs.

\end{itemize}

Our study provides valuable insights into the PDMF of star clusters in the solar neighborhood. The PDMF is heavily shaped by the internal and external dynamical evolution. The predominantly younger age range of our cluster sample emphasizes the need for observations of older and more massive open clusters that are located further away from the solar neighborhood, to refine star cluster models that include stellar mass black holes. The incompleteness limit from observations will need to be carefully considered when investigating distant old clusters.

\acknowledgments
{  We thank the anonymous referee for providing helpful comments and suggestions that helped to improve this paper.}
We thank Dr. Long Wang and {  Dr. Abbas Askar} for helpful discussions. 
Xiaoying Pang acknowledges the financial support of the National Natural Science Foundation of China through grants 12173029 and 12233013, and the research development fund of Xi'an Jiaotong-Liverpool University (RDF-18--02--32). 
M.B.N.K. acknowledges support from the National Natural Science Foundation of China {  (grant 11573004)}. 
Z.Y.\ acknowledges the support from the Jiangsu Funding Program for Excellent Postdoctoral Talent under grant number 2022ZB54, the National Natural Science Foundation of China under grant numbers 12203021, 12041305, and 12173016, and the Fundamental Research Funds for the Central Universities under grant number 0201/14380049.

This work made use of data from the European Space Agency (ESA) mission {\it Gaia} 
(\url{https://www.cosmos.esa.int/gaia}), processed by the {\it Gaia} Data Processing 
and Analysis Consortium (DPAC, \url{https://www.cosmos.esa.int/web/gaia/dpac/consortium}).

%This study also made use of the SIMBAD database and the VizieR catalogue access tool, both operated at CDS, Strasbourg, France.

%------------------

\software{  \texttt{Astropy} \citep{astropy2013,astropy2018,astropy2022}, 
            \texttt{SciPy} \citep{millman2011},
            %\texttt{TOPCAT} \citep{taylor2005}, 
            and 
            \textsc{StarGO} \citep{yuan2018}.
}
%-------------------
\clearpage

\appendix{}
\counterwithin{figure}{section}
\counterwithin{table}{section}

\section{Member Selection for NGC\,6991 with \textsc{StarGO}} \label{sec:NGC6991}

The \textsc{StarGO} \citep[][]{yuan2018} software utilizes the Self-Organizing Map (SOM) algorithm to project high-dimensional data onto a 100$\times$100 2D neural network, facilitating the identification of groupings in multi-dimensional parameter space. Each neuron in the network is assigned a random 5D weight vector corresponding to the input data dimensions: $X$, $Y$, $Z$, \pmra, and \pmdec. The training process involves sequentially presenting stars to the neural network, with \textsc{StarGO} updating the 5D weight vectors of the selected neuron and its neighbors to better match the input data. This iterative training, performed 400 times for convergence, results in neurons with similar 5D weight vectors forming patches in the 2D neural network, which represents member stars of the cluster. The field star contamination rate is estimated using the mock Gaia EDR\,3 catalog \citep{rybizki2020}. We restrict the membership of NGC\,6991 to a 5\% contamination rate. The resulting 100$\times$100 2D neural network for the cluster NGC\,6991 is presented in Figure~\ref{fig:ngc6991} (a). The blue patch corresponds to a 5\% membership contamination rate.

We display the member stars of NGC\,6991 in the {  CMD} (Figure~\ref{fig:ngc6991} (b)), which shows a clear locus of a main sequence and red giant stars. PARSEC isochrones \citep{bressan2012} are adopted to fit the age of NGC\,6991. This best-fitted isochrone of 1.4\,Gyr (black curve in Figure~\ref{fig:ngc6991} (b)) is used for the generation of the synthetic unresolved binary populations in this cluster (see Section~\ref{sec:BC}).

Panel (c) in Figure~\ref{fig:ngc6991} displays the 2D spatial distribution of member stars, revealing the presence of elongated tail structures. Such morphological tails are frequently accompanied with kinematic tails, as noted in previous studies \citep{li2021}. Correspondingly, the proper motion distribution of NGC\,6991 exhibits a conspicuous extended pattern in panel (d), indicating the disrupted state of this old cluster.

\begin{figure}[b!]
\centering
\includegraphics[angle=0, width=0.5\textwidth]{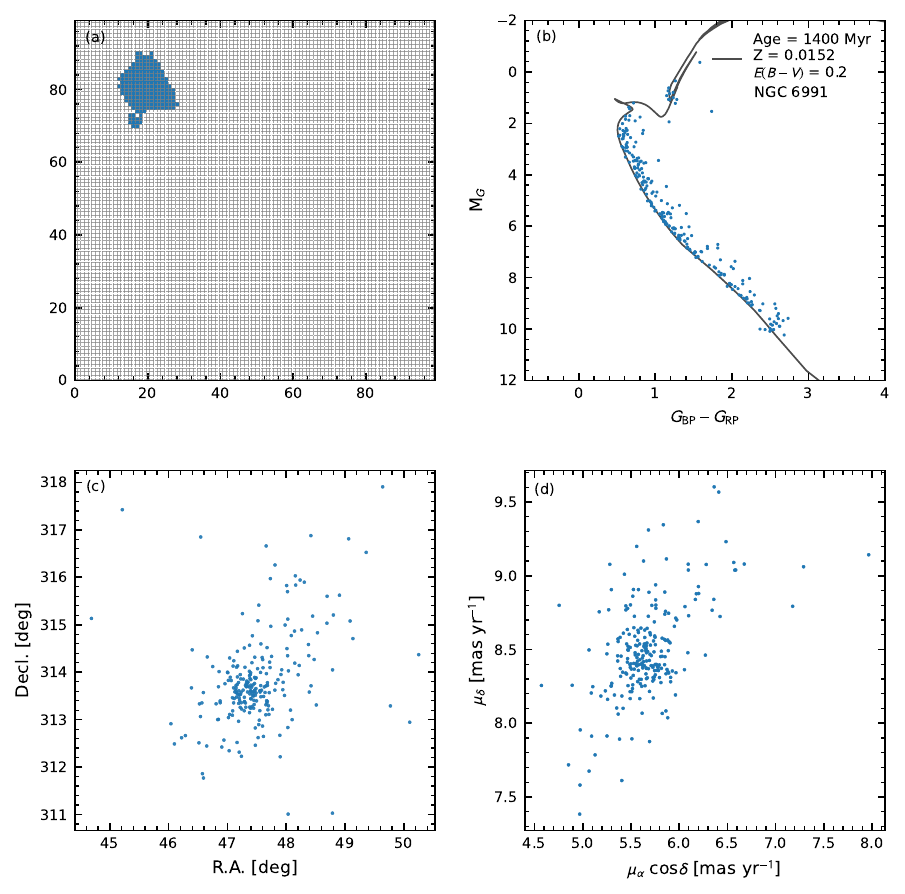}
    \caption{
 (a): A 100$\times$100 2D neural network generated by \textsc{StarGO} for NGC\,6991. Each grid is one neuron. Neurons corresponding to member stars (5\% contamination rate) formed a blue patch in this figure. (b): {  CMD} obtained from the Gaia DR\,3 absolute magnitude ${\rm M}_G$ for member stars in NGC\,6991. The PARSEC isochrones of the fitted age are indicated with the black solid curve, the fitted extinction and metallicity are also indicated in the upper right corner. (c): Spatial distribution of member stars (blue dots) selected by \textsc{StarGO}. (d) The proper motion vector plot for member stars. }
\label{fig:ngc6991}
\end{figure}

%%%%%%%%%%%%%%%%%%%%%%%%%%%%%%%%%%%%%%%%%%%%%%%%%%%%%%%
\newpage
\bibliography{main}

\begin{thebibliography}{}
\expandafter\ifx\csname natexlab\endcsname\relax\def\natexlab#1{#1}\fi
\providecommand{\url}[1]{\href{#1}{#1}}
\providecommand{\dodoi}[1]{doi:~\href{http://doi.org/#1}{\nolinkurl{#1}}}
\providecommand{\doeprint}[1]{\href{http://ascl.net/#1}{\nolinkurl{http://ascl.net/#1}}}
\providecommand{\doarXiv}[1]{\href{https://arxiv.org/abs/#1}{\nolinkurl{https://arxiv.org/abs/#1}}}

\bibitem[{{Abdurro'uf} {et~al.}(2022){Abdurro'uf}, {Accetta}, {Aerts}, {Silva
  Aguirre}, {Ahumada}, {Ajgaonkar}, {Filiz Ak}, {Alam}, {Allende Prieto},
  {Almeida}, {Anders}, {Anderson}, {Andrews}, {Anguiano}, {Aquino-Ort{\'\i}z},
  {Arag{\'o}n-Salamanca}, {Argudo-Fern{\'a}ndez}, {Ata}, {Aubert},
  {Avila-Reese}, {Badenes}, {Barb{\'a}}, {Barger}, {Barrera-Ballesteros},
  {Beaton}, {Beers}, {Belfiore}, {Bender}, {Bernardi}, {Bershady}, {Beutler},
  {Bidin}, {Bird}, {Bizyaev}, {Blanc}, {Blanton}, {Boardman}, {Bolton},
  {Boquien}, {Borissova}, {Bovy}, {Brandt}, {Brown}, {Brownstein}, {Brusa},
  {Buchner}, {Bundy}, {Burchett}, {Bureau}, {Burgasser}, {Cabang}, {Campbell},
  {Cappellari}, {Carlberg}, {Wanderley}, {Carrera}, {Cash}, {Chen}, {Chen},
  {Cherinka}, {Chiappini}, {Choi}, {Chojnowski}, {Chung}, {Clerc}, {Cohen},
  {Comerford}, {Comparat}, {da Costa}, {Covey}, {Crane}, {Cruz-Gonzalez},
  {Culhane}, {Cunha}, {Dai}, {Damke}, {Darling}, {Davidson}, {Davies},
  {Dawson}, {De Lee}, {Diamond-Stanic}, {Cano-D{\'\i}az}, {S{\'a}nchez},
  {Donor}, {Duckworth}, {Dwelly}, {Eisenstein}, {Elsworth}, {Emsellem},
  {Eracleous}, {Escoffier}, {Fan}, {Farr}, {Feng}, {Fern{\'a}ndez-Trincado},
  {Feuillet}, {Filipp}, {Fillingham}, {Frinchaboy}, {Fromenteau}, {Galbany},
  {Garc{\'\i}a}, {Garc{\'\i}a-Hern{\'a}ndez}, {Ge}, {Geisler}, {Gelfand},
  {G{\'e}ron}, {Gibson}, {Goddy}, {Godoy-Rivera}, {Grabowski}, {Green},
  {Greener}, {Grier}, {Griffith}, {Guo}, {Guy}, {Hadjara}, {Harding},
  {Hasselquist}, {Hayes}, {Hearty}, {Hern{\'a}ndez}, {Hill}, {Hogg},
  {Holtzman}, {Horta}, {Hsieh}, {Hsu}, {Hsu}, {Huber}, {Huertas-Company},
  {Hutchinson}, {Hwang}, {Ibarra-Medel}, {Chitham}, {Ilha}, {Imig}, {Jaekle},
  {Jayasinghe}, {Ji}, {Johnson}, {Jones}, {J{\"o}nsson}, {Katkov}, {Khalatyan},
  {Kinemuchi}, {Kisku}, {Knapen}, {Kneib}, {Kollmeier}, {Kong}, {Kounkel},
  {Kreckel}, {Krishnarao}, {Lacerna}, {Lane}, {Langgin}, {Lavender}, {Law},
  {Lazarz}, {Leung}, {Leung}, {Lewis}, {Li}, {Li}, {Lian}, {Liang}, {Lin},
  {Lin}, {Lin}, {Lintott}, {Long}, {Longa-Pe{\~n}a}, {L{\'o}pez-Cob{\'a}},
  {Lu}, {Lundgren}, {Luo}, {Mackereth}, {de la Macorra}, {Mahadevan},
  {Majewski}, {Manchado}, {Mandeville}, {Maraston}, {Margalef-Bentabol},
  {Masseron}, {Masters}, {Mathur}, {McDermid}, {Mckay}, {Merloni},
  {Merrifield}, {Meszaros}, {Miglio}, {Di Mille}, {Minniti}, {Minsley},
  {Monachesi}, {Moon}, {Mosser}, {Mulchaey}, {Muna}, {Mu{\~n}oz}, {Myers},
  {Myers}, {Nadathur}, {Nair}, {Nandra}, {Neumann}, {Newman}, {Nidever},
  {Nikakhtar}, {Nitschelm}, {O'Connell}, {Garma-Oehmichen}, {Luan Souza de
  Oliveira}, {Olney}, {Oravetz}, {Ortigoza-Urdaneta}, {Osorio}, {Otter},
  {Pace}, {Padilla}, {Pan}, {Pan}, {Parikh}, {Parker}, {Peirani}, {Pe{\~n}a
  Ram{\'\i}rez}, {Penny}, {Percival}, {Perez-Fournon}, {Pinsonneault},
  {Poidevin}, {Poovelil}, {Price-Whelan}, {B{\'a}rbara de Andrade Queiroz},
  {Raddick}, {Ray}, {Rembold}, {Riddle}, {Riffel}, {Riffel}, {Rix}, {Robin},
  {Rodr{\'\i}guez-Puebla}, {Roman-Lopes}, {Rom{\'a}n-Z{\'u}{\~n}iga}, {Rose},
  {Ross}, {Rossi}, {Rubin}, {Salvato}, {S{\'a}nchez}, {S{\'a}nchez-Gallego},
  {Sanderson}, {Santana Rojas}, {Sarceno}, {Sarmiento}, {Sayres}, {Sazonova},
  {Schaefer}, {Schiavon}, {Schlegel}, {Schneider}, {Schultheis}, {Schwope},
  {Serenelli}, {Serna}, {Shao}, {Shapiro}, {Sharma}, {Shen}, {Shetrone}, {Shu},
  {Simon}, {Skrutskie}, {Smethurst}, {Smith}, {Sobeck}, {Spoo}, {Sprague},
  {Stark}, {Stassun}, {Steinmetz}, {Stello}, {Stone-Martinez},
  {Storchi-Bergmann}, {Stringfellow}, {Stutz}, {Su}, {Taghizadeh-Popp},
  {Talbot}, {Tayar}, {Telles}, {Teske}, {Thakar}, {Theissen}, {Tkachenko},
  {Thomas}, {Tojeiro}, {Hernandez Toledo}, {Troup}, {Trump}, {Trussler},
  {Turner}, {Tuttle}, {Unda-Sanzana}, {V{\'a}zquez-Mata}, {Valentini},
  {Valenzuela}, {Vargas-Gonz{\'a}lez}, {Vargas-Maga{\~n}a}, {Alfaro},
  {Villanova}, {Vincenzo}, {Wake}, {Warfield}, {Washington}, {Weaver},
  {Weijmans}, {Weinberg}, {Weiss}, {Westfall}, {Wild}, {Wilde}, {Wilson},
  {Wilson}, {Wilson}, {Wolf}, {Wood-Vasey}, {Yan}, {Zamora}, {Zasowski},
  {Zhang}, {Zhao}, {Zheng}, {Zheng}, \& {Zhu}}]{abdurrouf2022ApJS..259...35A}
{Abdurro'uf}, {Accetta}, K., {Aerts}, C., {et~al.} 2022, \apjs, 259, 35,
  \dodoi{10.3847/1538-4365/ac4414}

\bibitem[{{Allison} {et~al.}(2009){Allison}, {Goodwin}, {Parker}, {de Grijs},
  {Portegies Zwart}, \& {Kouwenhoven}}]{allison2009}
{Allison}, R.~J., {Goodwin}, S.~P., {Parker}, R.~J., {et~al.} 2009, \apjl, 700,
  L99, \dodoi{10.1088/0004-637X/700/2/L99}

\bibitem[{{Aoyama} {et~al.}(2023){Aoyama}, {Ouchi}, \& {Harikane}}]{aoyama2023}
{Aoyama}, S., {Ouchi}, M., \& {Harikane}, Y. 2023, \apj, 946, 69,
  \dodoi{10.3847/1538-4357/acba87}

\bibitem[{{Astropy Collaboration} {et~al.}(2013){Astropy Collaboration},
  {Robitaille}, {Tollerud}, {Greenfield}, {Droettboom}, {Bray}, {Aldcroft},
  {Davis}, {Ginsburg}, {Price-Whelan}, {Kerzendorf}, {Conley}, {Crighton},
  {Barbary}, {Muna}, {Ferguson}, {Grollier}, {Parikh}, {Nair}, {Unther},
  {Deil}, {Woillez}, {Conseil}, {Kramer}, {Turner}, {Singer}, {Fox}, {Weaver},
  {Zabalza}, {Edwards}, {Azalee Bostroem}, {Burke}, {Casey}, {Crawford},
  {Dencheva}, {Ely}, {Jenness}, {Labrie}, {Lim}, {Pierfederici}, {Pontzen},
  {Ptak}, {Refsdal}, {Servillat}, \& {Streicher}}]{astropy2013}
{Astropy Collaboration}, {Robitaille}, T.~P., {Tollerud}, E.~J., {et~al.} 2013,
  \aap, 558, A33, \dodoi{10.1051/0004-6361/201322068}

\bibitem[{{Astropy Collaboration} {et~al.}(2018){Astropy Collaboration},
  {Price-Whelan}, {Sip{\H{o}}cz}, {G{\"u}nther}, {Lim}, {Crawford}, {Conseil},
  {Shupe}, {Craig}, {Dencheva}, {Ginsburg}, {Vand erPlas}, {Bradley},
  {P{\'e}rez-Su{\'a}rez}, {de Val-Borro}, {Aldcroft}, {Cruz}, {Robitaille},
  {Tollerud}, {Ardelean}, {Babej}, {Bach}, {Bachetti}, {Bakanov}, {Bamford},
  {Barentsen}, {Barmby}, {Baumbach}, {Berry}, {Biscani}, {Boquien}, {Bostroem},
  {Bouma}, {Brammer}, {Bray}, {Breytenbach}, {Buddelmeijer}, {Burke},
  {Calderone}, {Cano Rodr{\'\i}guez}, {Cara}, {Cardoso}, {Cheedella}, {Copin},
  {Corrales}, {Crichton}, {D'Avella}, {Deil}, {Depagne}, {Dietrich}, {Donath},
  {Droettboom}, {Earl}, {Erben}, {Fabbro}, {Ferreira}, {Finethy}, {Fox},
  {Garrison}, {Gibbons}, {Goldstein}, {Gommers}, {Greco}, {Greenfield},
  {Groener}, {Grollier}, {Hagen}, {Hirst}, {Homeier}, {Horton}, {Hosseinzadeh},
  {Hu}, {Hunkeler}, {Ivezi{\'c}}, {Jain}, {Jenness}, {Kanarek}, {Kendrew},
  {Kern}, {Kerzendorf}, {Khvalko}, {King}, {Kirkby}, {Kulkarni}, {Kumar},
  {Lee}, {Lenz}, {Littlefair}, {Ma}, {Macleod}, {Mastropietro}, {McCully},
  {Montagnac}, {Morris}, {Mueller}, {Mumford}, {Muna}, {Murphy}, {Nelson},
  {Nguyen}, {Ninan}, {N{\"o}the}, {Ogaz}, {Oh}, {Parejko}, {Parley}, {Pascual},
  {Patil}, {Patil}, {Plunkett}, {Prochaska}, {Rastogi}, {Reddy Janga},
  {Sabater}, {Sakurikar}, {Seifert}, {Sherbert}, {Sherwood-Taylor}, {Shih},
  {Sick}, {Silbiger}, {Singanamalla}, {Singer}, {Sladen}, {Sooley},
  {Sornarajah}, {Streicher}, {Teuben}, {Thomas}, {Tremblay}, {Turner},
  {Terr{\'o}n}, {van Kerkwijk}, {de la Vega}, {Watkins}, {Weaver}, {Whitmore},
  {Woillez}, {Zabalza}, \& {Astropy Contributors}}]{astropy2018}
{Astropy Collaboration}, {Price-Whelan}, A.~M., {Sip{\H{o}}cz}, B.~M., {et~al.}
  2018, \aj, 156, 123, \dodoi{10.3847/1538-3881/aabc4f}

\bibitem[{{Astropy Collaboration} {et~al.}(2022){Astropy Collaboration},
  {Price-Whelan}, {Lim}, {Earl}, {Starkman}, {Bradley}, {Shupe}, {Patil},
  {Corrales}, {Brasseur}, {N{"o}the}, {Donath}, {Tollerud}, {Morris},
  {Ginsburg}, {Vaher}, {Weaver}, {Tocknell}, {Jamieson}, {van Kerkwijk},
  {Robitaille}, {Merry}, {Bachetti}, {G{"u}nther}, {Aldcroft},
  {Alvarado-Montes}, {Archibald}, {B{'o}di}, {Bapat}, {Barentsen}, {Baz{'a}n},
  {Biswas}, {Boquien}, {Burke}, {Cara}, {Cara}, {Conroy}, {Conseil}, {Craig},
  {Cross}, {Cruz}, {D'Eugenio}, {Dencheva}, {Devillepoix}, {Dietrich},
  {Eigenbrot}, {Erben}, {Ferreira}, {Foreman-Mackey}, {Fox}, {Freij}, {Garg},
  {Geda}, {Glattly}, {Gondhalekar}, {Gordon}, {Grant}, {Greenfield}, {Groener},
  {Guest}, {Gurovich}, {Handberg}, {Hart}, {Hatfield-Dodds}, {Homeier},
  {Hosseinzadeh}, {Jenness}, {Jones}, {Joseph}, {Kalmbach}, {Karamehmetoglu},
  {Ka{l}uszy{'n}ski}, {Kelley}, {Kern}, {Kerzendorf}, {Koch}, {Kulumani},
  {Lee}, {Ly}, {Ma}, {MacBride}, {Maljaars}, {Muna}, {Murphy}, {Norman},
  {O'Steen}, {Oman}, {Pacifici}, {Pascual}, {Pascual-Granado}, {Patil},
  {Perren}, {Pickering}, {Rastogi}, {Roulston}, {Ryan}, {Rykoff}, {Sabater},
  {Sakurikar}, {Salgado}, {Sanghi}, {Saunders}, {Savchenko}, {Schwardt},
  {Seifert-Eckert}, {Shih}, {Jain}, {Shukla}, {Sick}, {Simpson},
  {Singanamalla}, {Singer}, {Singhal}, {Sinha}, {Sip{H{o}}cz}, {Spitler},
  {Stansby}, {Streicher}, {{{S}}umak}, {Swinbank}, {Taranu}, {Tewary},
  {Tremblay}, {Val-Borro}, {Van Kooten}, {Vasovi{'c}}, {Verma}, {de Miranda
  Cardoso}, {Williams}, {Wilson}, {Winkel}, {Wood-Vasey}, {Xue}, {Yoachim},
  {Zhang}, {Zonca}, \& {Astropy Project Contributors}}]{astropy2022}
{Astropy Collaboration}, {Price-Whelan}, A.~M., {Lim}, P.~L., {et~al.} 2022,
  apj, 935, 167, \dodoi{10.3847/1538-4357/ac7c74}

\bibitem[{Bastian {et~al.}(2010)Bastian, Covey, \&
  Meyer}]{bastian_universal_2010}
Bastian, N., Covey, K.~R., \& Meyer, M.~R. 2010, Annual Review of Astronomy and
  Astrophysics, 48, 339, \dodoi{10.1146/annurev-astro-082708-101642}

\bibitem[{Baumgardt {et~al.}(2023)Baumgardt, Hénault-Brunet, Dickson, \&
  Sollima}]{baumgardt_evidence_2023}
Baumgardt, H., Hénault-Brunet, V., Dickson, N., \& Sollima, A. 2023, Monthly
  Notices of the Royal Astronomical Society, 521, 3991,
  \dodoi{10.1093/mnras/stad631}

\bibitem[{Baumgardt \& Makino(2003)}]{baumgardt_dynamical_2003}
Baumgardt, H., \& Makino, J. 2003, Monthly Notices of the Royal Astronomical
  Society, 340, 227, \dodoi{10.1046/j.1365-8711.2003.06286.x}

\bibitem[{{Baumgardt} \& {Makino}(2003)}]{2003MNRAS.340..227B}
{Baumgardt}, H., \& {Makino}, J. 2003, \mnras, 340, 227,
  \dodoi{10.1046/j.1365-8711.2003.06286.x}

\bibitem[{{Belczynski} {et~al.}(2010){Belczynski}, {Bulik}, {Fryer}, {Ruiter},
  {Valsecchi}, {Vink}, \& {Hurley}}]{belczynski2010}
{Belczynski}, K., {Bulik}, T., {Fryer}, C.~L., {et~al.} 2010, \apj, 714, 1217,
  \dodoi{10.1088/0004-637X/714/2/1217}

\bibitem[{{Binney} \& {Tremaine}(2008)}]{binney2008}
{Binney}, J., \& {Tremaine}, S. 2008, {Galactic Dynamics: Second Edition}
  (Princeton University Press)

\bibitem[{{Breen} \& {Heggie}(2013)}]{Breen2013}
{Breen}, P.~G., \& {Heggie}, D.~C. 2013, \mnras, 432, 2779,
  \dodoi{10.1093/mnras/stt628}

\bibitem[{{Bressan} {et~al.}(2012){Bressan}, {Marigo}, {Girardi}, {Salasnich},
  {Dal Cero}, {Rubele}, \& {Nanni}}]{bressan2012}
{Bressan}, A., {Marigo}, P., {Girardi}, L., {et~al.} 2012, \mnras, 427, 127,
  \dodoi{10.1111/j.1365-2966.2012.21948.x}

\bibitem[{{Bromm} {et~al.}(2009){Bromm}, {Yoshida}, {Hernquist}, \&
  {McKee}}]{Bromm2009}
{Bromm}, V., {Yoshida}, N., {Hernquist}, L., \& {McKee}, C.~F. 2009, \nat, 459,
  49, \dodoi{10.1038/nature07990}

\bibitem[{{Calura} \& {Menci}(2009)}]{calura2009}
{Calura}, F., \& {Menci}, N. 2009, \mnras, 400, 1347,
  \dodoi{10.1111/j.1365-2966.2009.15440.x}

\bibitem[{{Cantat-Gaudin} {et~al.}(2020){Cantat-Gaudin}, {Anders},
  {Castro-Ginard}, {Jordi}, {Romero-Gomez}, {Soubiran}, {Casamiquela},
  {Tarricq}, {Moitinho}, {Vallenari}, {Bragaglia}, {Krone-Martins}, \&
  {Kounkel}}]{Cantat-Gaudin2020}
{Cantat-Gaudin}, T., {Anders}, F., {Castro-Ginard}, A., {et~al.} 2020, VizieR
  Online Data Catalog, J/A+A/640/A1

\bibitem[{{Chabrier}(2003)}]{chabrier2003}
{Chabrier}, G. 2003, \pasp, 115, 763, \dodoi{10.1086/376392}

\bibitem[{{Di Carlo} {et~al.}(2020){Di Carlo}, {Mapelli}, {Bouffanais},
  {Giacobbo}, {Santoliquido}, {Bressan}, {Spera}, \& {Haardt}}]{dicarlo2020}
{Di Carlo}, U.~N., {Mapelli}, M., {Bouffanais}, Y., {et~al.} 2020, \mnras, 497,
  1043, \dodoi{10.1093/mnras/staa1997}

\bibitem[{Fisher {et~al.}(2005)Fisher, Schröder, \& Smith}]{fisher_what_2005}
Fisher, J., Schröder, K.-P., \& Smith, R.~C. 2005, Monthly Notices of the
  Royal Astronomical Society, 361, 495,
  \dodoi{10.1111/j.1365-2966.2005.09193.x}

\bibitem[{{Gaia Collaboration} {et~al.}(2022){Gaia Collaboration}, Vallenari,
  Brown, Prusti, \& {et al.}}]{gaia_collaboration_gaia_2022}
{Gaia Collaboration}, Vallenari, A., Brown, A., Prusti, T., \& {et al.} 2022,
  Astronomy \& Astrophysics, \dodoi{10.1051/0004-6361/202243940}

\bibitem[{{Gaia Collaboration} {et~al.}(2021){Gaia Collaboration}, {Brown},
  {Vallenari}, {Prusti}, {de Bruijne}, {Babusiaux}, {Biermann}, {Creevey},
  {Evans}, {Eyer}, {Hutton}, {Jansen}, {Jordi}, {Klioner}, {Lammers},
  {Lindegren}, {Luri}, {Mignard}, {Panem}, {Pourbaix}, {Randich}, {Sartoretti},
  {Soubiran}, {Walton}, {Arenou}, {Bailer-Jones}, {Bastian}, {Cropper},
  {Drimmel}, {Katz}, {Lattanzi}, {van Leeuwen}, {Bakker}, {Cacciari},
  {Casta{\~n}eda}, {De Angeli}, {Ducourant}, {Fabricius}, {Fouesneau},
  {Fr{\'e}mat}, {Guerra}, {Guerrier}, {Guiraud}, {Jean-Antoine Piccolo},
  {Masana}, {Messineo}, {Mowlavi}, {Nicolas}, {Nienartowicz}, {Pailler},
  {Panuzzo}, {Riclet}, {Roux}, {Seabroke}, {Sordo}, {Tanga}, {Th{\'e}venin},
  {Gracia-Abril}, {Portell}, {Teyssier}, {Altmann}, {Andrae}, {Bellas-Velidis},
  {Benson}, {Berthier}, {Blomme}, {Brugaletta}, {Burgess}, {Busso}, {Carry},
  {Cellino}, {Cheek}, {Clementini}, {Damerdji}, {Davidson}, {Delchambre},
  {Dell'Oro}, {Fern{\'a}ndez-Hern{\'a}ndez}, {Galluccio}, {Garc{\'\i}a-Lario},
  {Garcia-Reinaldos}, {Gonz{\'a}lez-N{\'u}{\~n}ez}, {Gosset}, {Haigron},
  {Halbwachs}, {Hambly}, {Harrison}, {Hatzidimitriou}, {Heiter},
  {Hern{\'a}ndez}, {Hestroffer}, {Hodgkin}, {Holl}, {Jan{\ss}en}, {Jevardat de
  Fombelle}, {Jordan}, {Krone-Martins}, {Lanzafame}, {L{\"o}ffler}, {Lorca},
  {Manteiga}, {Marchal}, {Marrese}, {Moitinho}, {Mora}, {Muinonen}, {Osborne},
  {Pancino}, {Pauwels}, {Petit}, {Recio-Blanco}, {Richards}, {Riello},
  {Rimoldini}, {Robin}, {Roegiers}, {Rybizki}, {Sarro}, {Siopis}, {Smith},
  {Sozzetti}, {Ulla}, {Utrilla}, {van Leeuwen}, {van Reeven}, {Abbas}, {Abreu
  Aramburu}, {Accart}, {Aerts}, {Aguado}, {Ajaj}, {Altavilla}, {{\'A}lvarez},
  {{\'A}lvarez Cid-Fuentes}, {Alves}, {Anderson}, {Anglada Varela}, {Antoja},
  {Audard}, {Baines}, {Baker}, {Balaguer-N{\'u}{\~n}ez}, {Balbinot}, {Balog},
  {Barache}, {Barbato}, {Barros}, {Barstow}, {Bartolom{\'e}}, {Bassilana},
  {Bauchet}, {Baudesson-Stella}, {Becciani}, {Bellazzini}, {Bernet}, {Bertone},
  {Bianchi}, {Blanco-Cuaresma}, {Boch}, {Bombrun}, {Bossini}, {Bouquillon},
  {Bragaglia}, {Bramante}, {Breedt}, {Bressan}, {Brouillet}, {Bucciarelli},
  {Burlacu}, {Busonero}, {Butkevich}, {Buzzi}, {Caffau}, {Cancelliere},
  {C{\'a}novas}, {Cantat-Gaudin}, {Carballo}, {Carlucci}, {Carnerero},
  {Carrasco}, {Casamiquela}, {Castellani}, {Castro-Ginard}, {Castro Sampol},
  {Chaoul}, {Charlot}, {Chemin}, {Chiavassa}, {Cioni}, {Comoretto}, {Cooper},
  {Cornez}, {Cowell}, {Crifo}, {Crosta}, {Crowley}, {Dafonte}, {Dapergolas},
  {David}, {David}, {de Laverny}, {De Luise}, {De March}, {De Ridder}, {de
  Souza}, {de Teodoro}, {de Torres}, {del Peloso}, {del Pozo}, {Delbo},
  {Delgado}, {Delgado}, {Delisle}, {Di Matteo}, {Diakite}, {Diener},
  {Distefano}, {Dolding}, {Eappachen}, {Edvardsson}, {Enke}, {Esquej}, {Fabre},
  {Fabrizio}, {Faigler}, {Fedorets}, {Fernique}, {Fienga}, {Figueras},
  {Fouron}, {Fragkoudi}, {Fraile}, {Franke}, {Gai}, {Garabato},
  {Garcia-Gutierrez}, {Garc{\'\i}a-Torres}, {Garofalo}, {Gavras}, {Gerlach},
  {Geyer}, {Giacobbe}, {Gilmore}, {Girona}, {Giuffrida}, {Gomel}, {Gomez},
  {Gonzalez-Santamaria}, {Gonz{\'a}lez-Vidal}, {Granvik},
  {Guti{\'e}rrez-S{\'a}nchez}, {Guy}, {Hauser}, {Haywood}, {Helmi}, {Hidalgo},
  {Hilger}, {H{\l}adczuk}, {Hobbs}, {Holland}, {Huckle}, {Jasniewicz},
  {Jonker}, {Juaristi Campillo}, {Julbe}, {Karbevska}, {Kervella}, {Khanna},
  {Kochoska}, {Kontizas}, {Kordopatis}, {Korn}, {Kostrzewa-Rutkowska},
  {Kruszy{\'n}ska}, {Lambert}, {Lanza}, {Lasne}, {Le Campion}, {Le Fustec},
  {Lebreton}, {Lebzelter}, {Leccia}, {Leclerc}, {Lecoeur-Taibi}, {Liao},
  {Licata}, {Lindstr{\o}m}, {Lister}, {Livanou}, {Lobel}, {Madrero Pardo},
  {Managau}, {Mann}, {Marchant}, {Marconi}, {Marcos Santos}, {Marinoni},
  {Marocco}, {Marshall}, {Martin Polo}, {Mart{\'\i}n-Fleitas}, {Masip},
  {Massari}, {Mastrobuono-Battisti}, {Mazeh}, {McMillan}, {Messina},
  {Michalik}, {Millar}, {Mints}, {Molina}, {Molinaro}, {Moln{\'a}r},
  {Montegriffo}, {Mor}, {Morbidelli}, {Morel}, {Morris}, {Mulone}, {Munoz},
  {Muraveva}, {Murphy}, {Musella}, {Noval}, {Ord{\'e}novic}, {Orr{\`u}},
  {Osinde}, {Pagani}, {Pagano}, {Palaversa}, {Palicio}, {Panahi}, {Pawlak},
  {Pe{\~n}alosa Esteller}, {Penttil{\"a}}, {Piersimoni}, {Pineau}, {Plachy},
  {Plum}, {Poggio}, {Poretti}, {Poujoulet}, {Pr{\v{s}}a}, {Pulone}, {Racero},
  {Ragaini}, {Rainer}, {Raiteri}, {Rambaux}, {Ramos}, {Ramos-Lerate}, {Re
  Fiorentin}, {Regibo}, {Reyl{\'e}}, {Ripepi}, {Riva}, {Rixon}, {Robichon},
  {Robin}, {Roelens}, {Rohrbasser}, {Romero-G{\'o}mez}, {Rowell}, {Royer},
  {Rybicki}, {Sadowski}, {Sagrist{\`a} Sell{\'e}s}, {Sahlmann}, {Salgado},
  {Salguero}, {Samaras}, {Sanchez Gimenez}, {Sanna}, {Santove{\~n}a},
  {Sarasso}, {Schultheis}, {Sciacca}, {Segol}, {Segovia}, {S{\'e}gransan},
  {Semeux}, {Shahaf}, {Siddiqui}, {Siebert}, {Siltala}, {Slezak}, {Smart},
  {Solano}, {Solitro}, {Souami}, {Souchay}, {Spagna}, {Spoto}, {Steele},
  {Steidelm{\"u}ller}, {Stephenson}, {S{\"u}veges}, {Szabados}, {Szegedi-Elek},
  {Taris}, {Tauran}, {Taylor}, {Teixeira}, {Thuillot}, {Tonello}, {Torra},
  {Torra}, {Turon}, {Unger}, {Vaillant}, {van Dillen}, {Vanel}, {Vecchiato},
  {Viala}, {Vicente}, {Voutsinas}, {Weiler}, {Wevers}, {Wyrzykowski}, {Yoldas},
  {Yvard}, {Zhao}, {Zorec}, {Zucker}, {Zurbach}, \& {Zwitter}}]{gaia2021}
{Gaia Collaboration}, {Brown}, A.~G.~A., {Vallenari}, A., {et~al.} 2021, \aap,
  649, A1, \dodoi{10.1051/0004-6361/202039657}

\bibitem[{{Geha} {et~al.}(2013){Geha}, {Brown}, {Tumlinson}, {Kalirai},
  {Simon}, {Kirby}, {VandenBerg}, {Mu{\~n}oz}, {Avila}, {Guhathakurta}, \&
  {Ferguson}}]{geha2013}
{Geha}, M., {Brown}, T.~M., {Tumlinson}, J., {et~al.} 2013, \apj, 771, 29,
  \dodoi{10.1088/0004-637X/771/1/29}

\bibitem[{{Gennaro} {et~al.}(2018){Gennaro}, {Geha}, {Tchernyshyov}, {Brown},
  {Avila}, {Conroy}, {Mu{\~n}oz}, {Simon}, \& {Tumlinson}}]{gennaro2018}
{Gennaro}, M., {Geha}, M., {Tchernyshyov}, K., {et~al.} 2018, \apj, 863, 38,
  \dodoi{10.3847/1538-4357/aaceff}

\bibitem[{{Gouliermis} {et~al.}(2004){Gouliermis}, {Keller}, {Kontizas},
  {Kontizas}, \& {Bellas-Velidis}}]{2004A&A...416..137G}
{Gouliermis}, D., {Keller}, S.~C., {Kontizas}, M., {Kontizas}, E., \&
  {Bellas-Velidis}, I. 2004, \aap, 416, 137, \dodoi{10.1051/0004-6361:20031702}

\bibitem[{{Iyer} {et~al.}(2023){Iyer}, {Line}, {Muirhead}, {Fortney}, \&
  {Gharib-Nezhad}}]{Iyer2023}
{Iyer}, A.~R., {Line}, M.~R., {Muirhead}, P.~S., {Fortney}, J.~J., \&
  {Gharib-Nezhad}, E. 2023, \apj, 944, 41, \dodoi{10.3847/1538-4357/acabc2}

\bibitem[{{Je{\v{r}}{\'a}bkov{\'a}}
  {et~al.}(2018{\natexlab{a}}){Je{\v{r}}{\'a}bkov{\'a}}, {Hasani Zonoozi},
  {Kroupa}, {Beccari}, {Yan}, {Vazdekis}, \& {Zhang}}]{2018A&A...620A..39J}
{Je{\v{r}}{\'a}bkov{\'a}}, T., {Hasani Zonoozi}, A., {Kroupa}, P., {et~al.}
  2018{\natexlab{a}}, \aap, 620, A39, \dodoi{10.1051/0004-6361/201833055}

\bibitem[{{Je{\v{r}}{\'a}bkov{\'a}}
  {et~al.}(2018{\natexlab{b}}){Je{\v{r}}{\'a}bkov{\'a}}, {Hasani Zonoozi},
  {Kroupa}, {Beccari}, {Yan}, {Vazdekis}, \& {Zhang}}]{jerabkov2018}
---. 2018{\natexlab{b}}, \aap, 620, A39, \dodoi{10.1051/0004-6361/201833055}

\bibitem[{Kouwenhoven {et~al.}(2007)Kouwenhoven, Brown, Portegies~Zwart, \&
  Kaper}]{kouwenhoven_primordial_2007}
Kouwenhoven, M. B.~N., Brown, A. G.~A., Portegies~Zwart, S.~F., \& Kaper, L.
  2007, Astronomy \& Astrophysics, 474, 77, \dodoi{10.1051/0004-6361:20077719}

\bibitem[{{Kouwenhoven} {et~al.}(2010){Kouwenhoven}, {Goodwin}, {Parker},
  {Davies}, {Malmberg}, \& {Kroupa}}]{kouwenhoven2010}
{Kouwenhoven}, M.~B.~N., {Goodwin}, S.~P., {Parker}, R.~J., {et~al.} 2010,
  \mnras, 404, 1835, \dodoi{10.1111/j.1365-2966.2010.16399.x}

\bibitem[{{Kroupa}(1995{\natexlab{a}})}]{1995MNRAS.277.1491K}
{Kroupa}, P. 1995{\natexlab{a}}, \mnras, 277, 1491,
  \dodoi{10.1093/mnras/277.4.1491}

\bibitem[{{Kroupa}(1995{\natexlab{b}})}]{1995MNRAS.277.1507K}
---. 1995{\natexlab{b}}, \mnras, 277, 1507, \dodoi{10.1093/mnras/277.4.1507}

\bibitem[{{Kroupa}(2001)}]{kroupa2001}
---. 2001, \mnras, 322, 231, \dodoi{10.1046/j.1365-8711.2001.04022.x}

\bibitem[{{Kroupa} {et~al.}(1993){Kroupa}, {Tout}, \& {Gilmore}}]{Kroupa1993}
{Kroupa}, P., {Tout}, C.~A., \& {Gilmore}, G. 1993, \mnras, 262, 545,
  \dodoi{10.1093/mnras/262.3.545}

\bibitem[{{Kruijssen}(2012)}]{kruijssen2012}
{Kruijssen}, J.~M.~D. 2012, \mnras, 426, 3008,
  \dodoi{10.1111/j.1365-2966.2012.21923.x}

\bibitem[{{Krumholz}(2014)}]{krumholz2014}
{Krumholz}, M.~R. 2014, \physrep, 539, 49,
  \dodoi{10.1016/j.physrep.2014.02.001}

\bibitem[{{Lada} \& {Lada}(2003)}]{lada2003}
{Lada}, C.~J., \& {Lada}, E.~A. 2003, \araa, 41, 57,
  \dodoi{10.1146/annurev.astro.41.011802.094844}

\bibitem[{Lee {et~al.}(2020)Lee, Offner, Hennebelle, André, Zinnecker,
  Ballesteros-Paredes, Inutsuka, \& Kruijssen}]{lee_origin_2020}
Lee, Y.-N., Offner, S. S.~R., Hennebelle, P., {et~al.} 2020, Space Science
  Reviews, 216, 70, \dodoi{10.1007/s11214-020-00699-2}

\bibitem[{{Li} {et~al.}(2023){Li}, {Liu}, {Zhang}, {Tian}, {Fu}, {Li}, \&
  {Yan}}]{li2023Natur}
{Li}, J., {Liu}, C., {Zhang}, Z.-Y., {et~al.} 2023, \nat, 613, 460,
  \dodoi{10.1038/s41586-022-05488-1}

\bibitem[{Li {et~al.}(2023b)Li, Wong, Hogg, Rix, \& Chandra}]{li2023b}
Li, J., Wong, K. W.~K., Hogg, D.~W., Rix, H.-W., \& Chandra, V. 2023b,
  {AspGap}: {Augmented} {Stellar} {Parameters} and {Abundances} for 23 million
  {RGB} stars from {Gaia} {XP} low-resolution spectra,  arXiv.
\newblock \url{http://arxiv.org/abs/2309.14294}

\bibitem[{{Li} {et~al.}(2021){Li}, {Pang}, \& {Tang}}]{li2021}
{Li}, Y., {Pang}, X., \& {Tang}, S.-Y. 2021, Research Notes of the American
  Astronomical Society, 5, 173, \dodoi{10.3847/2515-5172/ac1688}

\bibitem[{{Liu} \& {Pang}(2019)}]{liu2019}
{Liu}, L., \& {Pang}, X. 2019, \apjs, 245, 32, \dodoi{10.3847/1538-4365/ab530a}

\bibitem[{{Magrini} {et~al.}(2023){Magrini}, {Viscasillas V{\'a}zquez},
  {Spina}, {Randich}, {Romano}, {Franciosini}, {Recio-Blanco}, {Nordlander},
  {D'Orazi}, {Baratella}, {Smiljanic}, {Dantas}, {Pasquini}, {Spitoni},
  {Casali}, {Van der Swaelmen}, {Bensby}, {Stonkute}, {Feltzing}, {Sacco},
  {Bragaglia}, {Pancino}, {Heiter}, {Biazzo}, {Gilmore}, {Bergemann},
  {Tautvai{\v{s}}ien{\.{e}}}, {Worley}, {Hourihane}, {Gonneau}, \&
  {Morbidelli}}]{Magrini2023}
{Magrini}, L., {Viscasillas V{\'a}zquez}, C., {Spina}, L., {et~al.} 2023, \aap,
  669, A119, \dodoi{10.1051/0004-6361/202244957}

\bibitem[{{Ma{\'\i}z Apell{\'a}niz} \& {{\'U}beda}(2005)}]{maiz2005}
{Ma{\'\i}z Apell{\'a}niz}, J., \& {{\'U}beda}, L. 2005, \apj, 629, 873,
  \dodoi{10.1086/431458}

\bibitem[{{Mart{\'\i}n-Navarro} {et~al.}(2015){Mart{\'\i}n-Navarro},
  {Vazdekis}, {La Barbera}, {Falc{\'o}n-Barroso}, {Lyubenova}, {van de Ven},
  {Ferreras}, {S{\'a}nchez}, {Trager}, {Garc{\'\i}a-Benito}, {Mast}, {Mendoza},
  {S{\'a}nchez-Bl{\'a}zquez}, {Gonz{\'a}lez Delgado}, {Walcher}, \& {CALIFA
  Team}}]{2015ApJ...806L..31M}
{Mart{\'\i}n-Navarro}, I., {Vazdekis}, A., {La Barbera}, F., {et~al.} 2015,
  \apjl, 806, L31, \dodoi{10.1088/2041-8205/806/2/L31}

\bibitem[{{McMillan} {et~al.}(2007){McMillan}, {Vesperini}, \& {Portegies
  Zwart}}]{mcmillan2007}
{McMillan}, S. L.~W., {Vesperini}, E., \& {Portegies Zwart}, S.~F. 2007, \apjl,
  655, L45, \dodoi{10.1086/511763}

\bibitem[{{Millman} \& {Aivazis}(2011)}]{millman2011}
{Millman}, K.~J., \& {Aivazis}, M. 2011, Computing in Science and Engineering,
  13, 9, \dodoi{10.1109/MCSE.2011.36}

\bibitem[{Minchev {et~al.}(2009)Minchev, Quillen, Williams, Freeman, Nordhaus,
  Siebert, \& Bienaymé}]{minchev_is_2009}
Minchev, I., Quillen, A.~C., Williams, M., {et~al.} 2009, Monthly Notices of
  the Royal Astronomical Society: Letters, 396, L56,
  \dodoi{10.1111/j.1745-3933.2009.00661.x}

\bibitem[{{Mor} {et~al.}(2019){Mor}, {Robin}, {Figueras}, {Roca-F{\`a}brega},
  \& {Luri}}]{2019A&A...624L...1M}
{Mor}, R., {Robin}, A.~C., {Figueras}, F., {Roca-F{\`a}brega}, S., \& {Luri},
  X. 2019, \aap, 624, L1, \dodoi{10.1051/0004-6361/201935105}

\bibitem[{Offner {et~al.}(2022)Offner, Moe, Kratter, Sadavoy, Jensen, \&
  Tobin}]{offner_origin_2022}
Offner, S. S.~R., Moe, M., Kratter, K.~M., {et~al.} 2022, The {Origin} and
  {Evolution} of {Multiple} {Star} {Systems},  arXiv.
\newblock \url{http://arxiv.org/abs/2203.10066}

\bibitem[{{Oh} \& {Kroupa}(2018)}]{oh2018}
{Oh}, S., \& {Kroupa}, P. 2018, \mnras, 481, 153, \dodoi{10.1093/mnras/sty2245}

\bibitem[{{Oh} {et~al.}(2015){Oh}, {Kroupa}, \& {Pflamm-Altenburg}}]{oh2015}
{Oh}, S., {Kroupa}, P., \& {Pflamm-Altenburg}, J. 2015, \apj, 805, 92,
  \dodoi{10.1088/0004-637X/805/2/92}

\bibitem[{{Pang} {et~al.}(2013){Pang}, {Grebel}, {Allison}, {Goodwin},
  {Altmann}, {Harbeck}, {Moffat}, \& {Drissen}}]{pang2013}
{Pang}, X., {Grebel}, E.~K., {Allison}, R.~J., {et~al.} 2013, \apj, 764, 73,
  \dodoi{10.1088/0004-637X/764/1/73}

\bibitem[{{Pang} {et~al.}(2021{\natexlab{a}}){Pang}, {Li}, {Yu}, {Tang},
  {Dinnbier}, {Kroupa}, {Pasquato}, \& {Kouwenhoven}}]{pang2021a}
{Pang}, X., {Li}, Y., {Yu}, Z., {et~al.} 2021{\natexlab{a}}, \apj, 912, 162,
  \dodoi{10.3847/1538-4357/abeaac}

\bibitem[{{Pang} {et~al.}(2022{\natexlab{a}}){Pang}, {Shu}, {Wang}, \&
  {Kouwenhoven}}]{pang2022b}
{Pang}, X., {Shu}, Q., {Wang}, L., \& {Kouwenhoven}, M.~B.~N.
  2022{\natexlab{a}}, Research in Astronomy and Astrophysics, 22, 095015,
  \dodoi{10.1088/1674-4527/ac7f0f}

\bibitem[{{Pang} {et~al.}(2021{\natexlab{b}}){Pang}, {Yu}, {Tang}, {Hong},
  {Yuan}, {Pasquato}, \& {Kouwenhoven}}]{pang2021b}
{Pang}, X., {Yu}, Z., {Tang}, S.-Y., {et~al.} 2021{\natexlab{b}}, \apj, 923,
  20, \dodoi{10.3847/1538-4357/ac2838}

\bibitem[{{Pang} {et~al.}(2022{\natexlab{b}}){Pang}, {Tang}, {Li}, {Yu},
  {Wang}, {Li}, {Li}, {Wang}, {Wang}, {Zhang}, {Pasquato}, \&
  {Kouwenhoven}}]{pang2022a}
{Pang}, X., {Tang}, S.-Y., {Li}, Y., {et~al.} 2022{\natexlab{b}}, \apj, 931,
  156, \dodoi{10.3847/1538-4357/ac674e}

\bibitem[{{Pang} {et~al.}(2022{\natexlab{c}}){Pang}, {Li}, {Tang}, {Wang},
  {Wang}, {Li}, {Wang}, {Kouwenhoven}, \& {Pasquato}}]{pang2022c}
{Pang}, X., {Li}, Y., {Tang}, S.-Y., {et~al.} 2022{\natexlab{c}}, \apjl, 937,
  L7, \dodoi{10.3847/2041-8213/ac8e68}

\bibitem[{{Pang} {et~al.}(2023){Pang}, {Wang}, {Tang}, {Rui}, {Bai}, {Li},
  {Feng}, {Kouwenhoven}, {Chen}, \& {Chuang}}]{pang2023}
{Pang}, X., {Wang}, Y., {Tang}, S.-Y., {et~al.} 2023, \aj, 166, 110,
  \dodoi{10.3847/1538-3881/ace76c}

\bibitem[{{Portegies Zwart} {et~al.}(2010){Portegies Zwart}, {McMillan}, \&
  {Gieles}}]{portegies2010}
{Portegies Zwart}, S.~F., {McMillan}, S. L.~W., \& {Gieles}, M. 2010, \araa,
  48, 431, \dodoi{10.1146/annurev-astro-081309-130834}

\bibitem[{{Reid} {et~al.}(2019){Reid}, {Menten}, {Brunthaler}, {Zheng}, {Dame},
  {Xu}, {Li}, {Sakai}, {Wu}, {Immer}, {Zhang}, {Sanna}, {Moscadelli}, {Rygl},
  {Bartkiewicz}, {Hu}, {Quiroga-Nu{\~n}ez}, \& {van Langevelde}}]{reid2019}
{Reid}, M.~J., {Menten}, K.~M., {Brunthaler}, A., {et~al.} 2019, \apj, 885,
  131, \dodoi{10.3847/1538-4357/ab4a11}

\bibitem[{{Riello} {et~al.}(2021){Riello}, {De Angeli}, {Evans}, {Montegriffo},
  {Carrasco}, {Busso}, {Palaversa}, {Burgess}, {Diener}, {Davidson}, {Rowell},
  {Fabricius}, {Jordi}, {Bellazzini}, {Pancino}, {Harrison}, {Cacciari}, {van
  Leeuwen}, {Hambly}, {Hodgkin}, {Osborne}, {Altavilla}, {Barstow}, {Brown},
  {Castellani}, {Cowell}, {De Luise}, {Gilmore}, {Giuffrida}, {Hidalgo},
  {Holland}, {Marinoni}, {Pagani}, {Piersimoni}, {Pulone}, {Ragaini}, {Rainer},
  {Richards}, {Sanna}, {Walton}, {Weiler}, \& {Yoldas}}]{riello2021}
{Riello}, M., {De Angeli}, F., {Evans}, D.~W., {et~al.} 2021, \aap, 649, A3,
  \dodoi{10.1051/0004-6361/202039587}

\bibitem[{{Rybizki} {et~al.}(2020){Rybizki}, {Demleitner}, {Bailer-Jones},
  {Tio}, {Cantat-Gaudin}, {Fouesneau}, {Chen}, {Andrae}, {Girardi}, \&
  {Sharma}}]{rybizki2020}
{Rybizki}, J., {Demleitner}, M., {Bailer-Jones}, C., {et~al.} 2020, \pasp, 132,
  074501, \dodoi{10.1088/1538-3873/ab8cb0}

\bibitem[{{Salpeter}(1955)}]{salpeter1955}
{Salpeter}, E.~E. 1955, \apj, 121, 161, \dodoi{10.1086/145971}

\bibitem[{{Sharda} \& {Krumholz}(2022)}]{2022MNRAS.509.1959S}
{Sharda}, P., \& {Krumholz}, M.~R. 2022, \mnras, 509, 1959,
  \dodoi{10.1093/mnras/stab2921}

\bibitem[{{Skillman}(2008)}]{2008IAUS..255..285S}
{Skillman}, E.~D. 2008, in Low-Metallicity Star Formation: From the First Stars
  to Dwarf Galaxies, ed. L.~K. {Hunt}, S.~C. {Madden}, \& R.~{Schneider}, Vol.
  255, 285--296, \dodoi{10.1017/S1743921308024964}

\bibitem[{{Smith}(2020)}]{2020ARA&A..58..577S}
{Smith}, R.~J. 2020, \araa, 58, 577,
  \dodoi{10.1146/annurev-astro-032620-020217}

\bibitem[{{Spera} {et~al.}(2019){Spera}, {Mapelli}, {Giacobbo}, {Trani},
  {Bressan}, \& {Costa}}]{spera2019}
{Spera}, M., {Mapelli}, M., {Giacobbo}, N., {et~al.} 2019, \mnras, 485, 889,
  \dodoi{10.1093/mnras/stz359}

\bibitem[{{Thies} {et~al.}(2015){Thies}, {Pflamm-Altenburg}, {Kroupa}, \&
  {Marks}}]{2015ApJ...800...72T}
{Thies}, I., {Pflamm-Altenburg}, J., {Kroupa}, P., \& {Marks}, M. 2015, \apj,
  800, 72, \dodoi{10.1088/0004-637X/800/1/72}

\bibitem[{Torniamenti {et~al.}(2023)Torniamenti, Gieles, Penoyre, Jerabkova,
  Wang, \& Anders}]{torniamenti_stellar-mass_2023}
Torniamenti, S., Gieles, M., Penoyre, Z., {et~al.} 2023, Monthly Notices of the
  Royal Astronomical Society, 524, 1965, \dodoi{10.1093/mnras/stad1925}

\bibitem[{{Vesperini} \& {Heggie}(1997)}]{vesperini1997}
{Vesperini}, E., \& {Heggie}, D.~C. 1997, \mnras, 289, 898,
  \dodoi{10.1093/mnras/289.4.898}

\bibitem[{{Weidner} {et~al.}(2010){Weidner}, {Kroupa}, \&
  {Bonnell}}]{2010MNRAS.401..275W}
{Weidner}, C., {Kroupa}, P., \& {Bonnell}, I.~A.~D. 2010, \mnras, 401, 275,
  \dodoi{10.1111/j.1365-2966.2009.15633.x}

\bibitem[{Yan {et~al.}(2017)Yan, Jerabkova, \& Kroupa}]{yan_optimally_2017}
Yan, Z., Jerabkova, T., \& Kroupa, P. 2017, Astronomy \& Astrophysics, 607,
  A126, \dodoi{10.1051/0004-6361/201730987}

\bibitem[{{Yan} {et~al.}(2020){Yan}, {Jerabkova}, \&
  {Kroupa}}]{2020A&A...637A..68Y}
{Yan}, Z., {Jerabkova}, T., \& {Kroupa}, P. 2020, \aap, 637, A68,
  \dodoi{10.1051/0004-6361/202037567}

\bibitem[{{Yan} {et~al.}(2023{\natexlab{a}}){Yan}, {Jerabkova}, \&
  {Kroupa}}]{yan2023}
---. 2023{\natexlab{a}}, \aap, 670, A151, \dodoi{10.1051/0004-6361/202244919}

\bibitem[{{Yan} {et~al.}(2023{\natexlab{b}}){Yan}, {Jerabkova}, \&
  {Kroupa}}]{2023A&A...670A.151Y}
---. 2023{\natexlab{b}}, \aap, 670, A151, \dodoi{10.1051/0004-6361/202244919}

\bibitem[{{Yan} {et~al.}(2021){Yan}, {Je{\v{r}}{\'a}bkov{\'a}}, \&
  {Kroupa}}]{2021A&A...655A..19Y}
{Yan}, Z., {Je{\v{r}}{\'a}bkov{\'a}}, T., \& {Kroupa}, P. 2021, \aap, 655, A19,
  \dodoi{10.1051/0004-6361/202140683}

\bibitem[{{Yang} {et~al.}(2024){Yang}, {Bird}, {Li}, {Tian}, {Qiu}, {Li}, {Li},
  {Liu}, {Zhang}, {Zhang}, \& {Chen}}]{yang2024}
{Yang}, X.-M., {Bird}, S.~A., {Li}, J., {et~al.} 2024, \mnras,
  \dodoi{10.1093/mnras/stae540}

\bibitem[{{Yuan} {et~al.}(2018){Yuan}, {Chang}, {Banerjee}, {Han}, {Kang}, \&
  {Smith}}]{yuan2018}
{Yuan}, Z., {Chang}, J., {Banerjee}, P., {et~al.} 2018, \apj, 863, 26,
  \dodoi{10.3847/1538-4357/aacd0d}

\end{thebibliography}
\bibliographystyle{aasjournal}
%%%%%%%%%%%%%%%%%%%%%%%%%%%%%%%%%%%%%%%%%%%%%%%%%%%%%%%
\end{document}